\pdfoutput=1

\documentclass[11pt,twoside,a4paper,cmspaper,final,collab]{cms-tdr}
\def\svnVersion{101242:102043}\def\cmsCernNo{2012-028}\def\cmsCernDate{\today}\def\cmsMessage{Submitted to the Journal of High Energy Physics}

\begin{document}\cmsNoteHeader{HIG-11-027}

\hyphenation{had-ron-i-za-tion}
\hyphenation{cal-or-i-me-ter}
\hyphenation{de-vices}

\RCS$Revision: 102043 $
\RCS$HeadURL: svn+ssh://alverson@svn.cern.ch/reps/tdr2/papers/HIG-11-027/trunk/HIG-11-027.tex $
\RCS$Id: HIG-11-027.tex 102043 2012-02-07 13:27:05Z alverson $
\cmsNoteHeader{HIG-11-006} 
\renewcommand{\MET}{\ensuremath{{E}^{\mathrm{miss}}_{\!\mathrm{T}}}\xspace}
\newcommand{\pp}{\ensuremath{\Pp\Pp}}%
\newcommand{\Wo}{\ensuremath{\PW}}%
\newcommand{\Wp}{\ensuremath{\PWp}}%
\newcommand{\Wm}{\ensuremath{\PWm}}%
\newcommand{\Zo}{\ensuremath{\cPZ}}%
\newcommand{\Ho}{\ensuremath{\PH}}%
\newcommand{\ra}{\ensuremath{\rightarrow}}%
\newcommand{\MN}{\ensuremath{\Pgm\cPgn}}%
\renewcommand{\LL}{\ensuremath{\ell^-\ell^+}}%
\renewcommand{\EE}{\ensuremath{\Pem\Pep}}%
\renewcommand{\TT}{\ensuremath{\ttbar}}%
\newcommand{\EN}{\ensuremath{\Pge\cPgn}}%
\newcommand{\LN}{\ensuremath{\ell\cPgn}}%
\newcommand{\MW}{\ensuremath{m_\Wo}}%
\newcommand{\MZ}{\ensuremath{m_\Zo}}%
\newcommand{\MT}{\ensuremath{M_\mathrm{T}}}%
\newcommand{\MLL}{\ensuremath{m_{\ell\ell}}}%
\newcommand{\nunubar}{\ensuremath{\cPgn\cPagn}}%
\newcommand{\qqbar}{\cPq\cPaq}%
\newcommand{\HZZ}{\ensuremath{\Ho\ra\Zo\Zo}}%
\newcommand{\HZZllqq}{\ensuremath{\HZZ^{(*)}\ra\qqbar\,\LL}}%
\newcommand{\Hllqq}{\ensuremath{\Ho\ra\LL\qqbar}}%
\newcommand{\mZ}{\ensuremath{m_{\Zo}}}%
\newcommand{\mW}{\ensuremath{m_{\Wo}}}%
\newcommand{\mH}{\ensuremath{m_{\Ho}}}%
\newcommand{\mZZ}{\ensuremath{m_{\Zo\Zo}}}%
\newcommand{\mll}{\ensuremath{m_{\ell\ell}}}%
\newcommand{\mjj}{\ensuremath{m_{jj}}}%
\newcommand{\GeVnn}{\ensuremath{{\,\text{Ge\hspace{-.08em}V\hspace{-0.16em}}}}\xspace}
\newcommand{\GeVcnn}{\ensuremath{{\,\text{Ge\hspace{-.08em}V\hspace{-0.16em}}}}\xspace}
\newcommand{\GeVsnn}{\ensuremath{{\text{Ge\hspace{-.08em}V\hspace{-0.16em}}}}\xspace}

\title{Search for a Higgs boson in the decay channel
$\PH\ra\cPZ\cPZ^{(*)}\ra\qqbar\LL$ in pp collisions at $\sqrt{s}$ = 7 TeV}

\date{\today}

\abstract{
A search for the standard model Higgs boson decaying into two
\cPZ\ bosons with subsequent decay into a final state containing two quark jets
and two leptons, $\PH\ra\cPZ\cPZ^{(*)}\ra\qqbar\LL$
is presented. Results are based on data corresponding to an integrated luminosity
of 4.6\fbinv of proton-proton collisions at $\sqrt{s}=7\TeV$,
collected with the CMS detector at the LHC.
In order to discriminate between signal and background events, kinematic and topological
quantities, including the angular spin correlations of the decay products, are employed.
Events are further classified according to the probability of the jets to originate from quarks of
light or heavy flavor or from gluons. No evidence for the Higgs boson is found,
and upper limits on its production cross section are determined for a Higgs boson
of mass between 130 and 600\GeV.
}

\hypersetup{%
pdfauthor={CMS Collaboration},%
pdftitle={Search for a Higgs boson in the decay channel H to ZZ(*) to q qbar l-l+ in pp collisions at sqrt(s) = 7 TeV},%
pdfsubject={CMS},%
pdfkeywords={CMS, physics, Higgs}}

\maketitle 

\section{Introduction}
An important goal of experiments at the Large Hadron Collider (LHC)~\cite{lhc-paper}
is to study the mechanism of electroweak symmetry breaking through which the weak $\Wo$ and $\Zo$
bosons acquire mass while the photon, \Pgg, remains massless. Within the standard model~(SM)
\cite{Glashow1961579, PhysRevLett.19.1264, Salam1968} of particle physics it is postulated that
the Higgs field provides the mechanism of electroweak symmetry
breaking~\cite{Englert:1964et, Higgs:1964ia,  Higgs:1964pj, Guralnik:1964eu, Higgs:1966ev, Kibble:1967sv}.
This model also predicts that the Higgs field would give rise to a spin-zero Higgs boson (H)
with quantum numbers of the vacuum, $J^{PC}=0^{++}$.
Limits set by the experiments at LEP~\cite{LEPHiggsLimit} and the Tevatron~\cite{TevHiggsLimit}
leave a wide range of allowed Higgs boson masses
$\mH > 114.4\GeV$ and $\mH \notin [162, 166]\GeV$ at 95\% confidence level (CL).
Recently, further limits were set by the ATLAS experiment~\cite{Aad:2011uq,ATLAS:2011af,ATLAS:2011aa}
at the LHC: $\mH \notin [145,206], [214,224],$ and $[340,450] \GeV$.
Indirect measurements~\cite{LEP:2010zz} suggest that the mass of a SM Higgs boson
would most likely fall below 158\GeV at 95\% CL.

At the LHC, within the SM, Higgs bosons are primarily produced by gluon fusion
(\cPg\cPg)~\cite{Dawson:1990zj,Spira:1995rr,Harlander:2002wh,Anastasiou:2002yz,Ravindran:2003um,Catani:2003zt,
Aglietti:2004nj,Actis:2008ug,Anastasiou:2008tj,deFlorian:2009hc} with an additional small
contribution due to weak vector boson fusion (VBF)~\cite{Ciccolini:2007jr,Ciccolini:2007ec,
Figy:2003nv,Arnold:2008rz,Bolzoni:2010xr,Figy:2010ct} and smaller contributions from other processes.
The decay of a Higgs boson to two light fermions is highly
suppressed~\cite{Djouadi:1997yw,hdecay2,Bredenstein:2006rh,Bredenstein:2006ha}.
Decay channels of the SM Higgs boson with two gauge bosons
in the final state provide the greatest discovery potential at the LHC.
For a Higgs boson mass $\mH < 2\mW$ those final states
contain two photons or two weak bosons, $\Zo\Zo^*$ or $\Wo\Wo^*$,
where in each case one of the gauge bosons is off mass shell.
For $\mH \ge 2\mW$, the main final states are those with two on-mass-shell weak bosons:
$\Wp\Wm$ for $2 \mW \le \mH < 2 \mZ$, and additionally $\Zo\Zo$ for $\mH \ge 2\mZ$.

In this Letter we present a search for a SM-like Higgs boson decaying
via two $\Zo$ bosons, one of which could be off mass shell,
with a subsequent decay into two quark jets and two leptons, $\HZZllqq$.
Constraints on the rate of the Higgs boson production and decay are presented as a function
of mass and interpretations are given in two scenarios: SM and a model with four generations
of fermions~\cite{Schmidt:2009kk,Li:2010fu,Anastasiou:2011pi, Denner:2011vt, Passarino:2011kv}.
The branching fraction of this decay channel is about 20 times higher
than that of $\HZZ^{(*)}\ra\LL\LL$. Inclusion of this semileptonic final state
in the search for the Higgs boson leads to improved sensitivity at higher masses,
where kinematic requirements can effectively suppress background.
In the low mass region with leptonically decaying off-mass-shell $\Zo$ bosons,
we can achieve effective background suppression by constraining the two
jets to the known $\Zo$ boson mass $\mZ$~\cite{Nakamura:2010zzi}.
The search is performed with a sample of proton-proton collisions
at a center-of-mass energy $\sqrt{s}$ = 7\TeV corresponding to an integrated luminosity
${\cal L}=(4.6\pm0.2)\fbinv$ recorded by the Compact Muon Solenoid (CMS)
experiment~\cite{CMS:2010} at the LHC during 2011.

\section{Event Reconstruction}
\label{sec:reco}
\begin{figure}[t!]
\begin{center}
\centerline{
\includegraphics[width=0.6\linewidth]{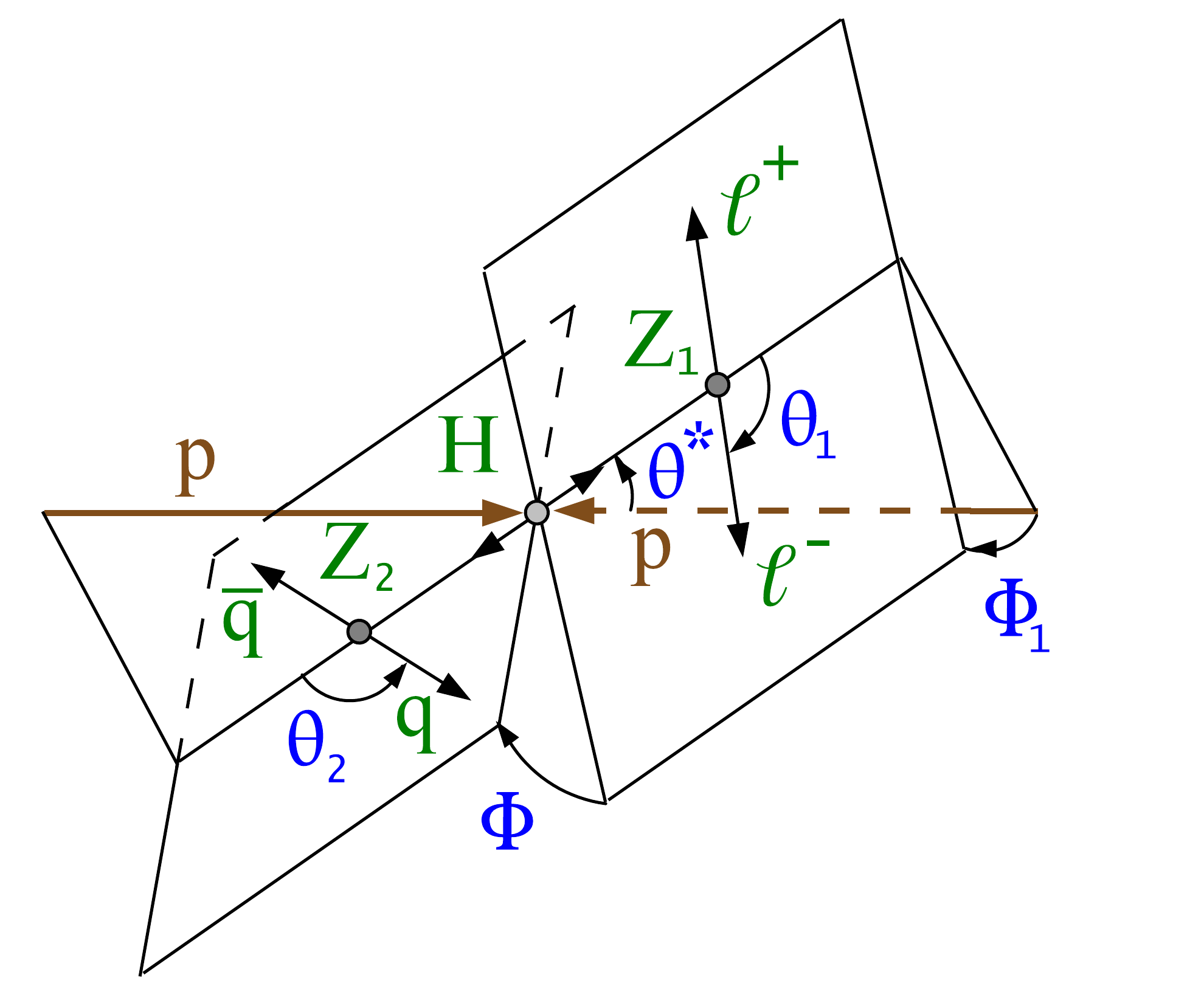}
}
\caption{
Diagram describing the process $\pp\to\Ho+\mathrm{X}\to\Zo\Zo^{(*)}+\mathrm{X}\to\qqbar\,\LL+\mathrm{X}$
in terms of the angles $(\theta^*, \Phi_1, \theta_1, \theta_2, \Phi)$ defined in the parent
particle rest frames ($\Ho$ or~$\Zo$), where $\mathrm{X}$ indicates other products of the $\pp$
collision not shown on the diagram~\cite{Gao:2010qx}.
}
\label{fig:angles-HZZ2l2q}
\end{center}
\end{figure}

We search for a fully reconstructed decay chain of the Higgs boson
$\HZZllqq$,  see figure~\ref{fig:angles-HZZ2l2q},
where the charged leptons $\ell^\pm$ are either muons or
electrons and the quarks are identified as jets in the CMS detector.
The search is optimized separately for two ranges of the reconstructed mass,
$125<\mZZ<170\GeV$ (low-mass)
and $183<\mZZ<800\GeV$ (high-mass),
corresponding to the $\HZZ^*$ and $\HZZ$ analyses, respectively.
The intermediate mass range between $2\mW < \mH < 2\mZ$ has reduced
sensitivity because of the small branching fraction for $\HZZ$
and is not included in the analysis.

A detailed description of the CMS detector can be found in ref.~\cite{CMS:2010}.
In the cylindrical coordinate system of CMS, $\phi$ is the azimuthal angle and
the pseudorapidity ($\eta$) is defined as $\eta = - \ln[\tan(\theta/2)]$, where $\theta$
is the polar angle with respect to the counterclockwise beam direction.
The central feature of the CMS detector is a 3.8\unit{T} superconducting solenoid of 6\unit{m} internal diameter.
Within the field volume are the silicon tracker,
the crystal electromagnetic calorimeter (ECAL), and the brass-scintillator hadron calorimeter (HCAL).
The muon system is installed outside the solenoid and embedded in the steel return yoke.
The CMS tracker consists of  silicon pixel and silicon strip detector modules,
covering the pseudorapidity range $|\eta|< 2.5$. The ECAL consists of lead tungstate crystals,
which provide coverage for pseudorapidity $\vert \eta \vert< 1.5$ in
the central barrel region and $1.5 <\vert \eta \vert < 3.0$ in the two forward endcap regions.
The HCAL consists of a set of sampling
calorimeters which utilize alternating layers of brass as absorber and plastic scintillator as active material.
The muon system includes barrel drift tubes covering the pseudorapidity range $|\eta|< 1.2$, endcap
cathode strip chambers ($0.9< |\eta|< 2.5$), and resistive plate chambers ($|\eta|< 1.6$).

Although the main sources of background are estimated from data, Monte Carlo (MC) simulations
are used to develop and validate the methods used in the analysis. Background samples are generated using
either \MADGRAPH4.4.12~\cite{MADGRAPH} (inclusive $\Zo$ and top-quark production),
\ALPGEN2.13~\cite{ALPGEN} (inclusive $\Zo$ production),
\POWHEG~\cite{Nason:2004,Frixione:2007,Alioli:2008} (top-quark production),
or \PYTHIA6.4.22~\cite{PYTHIA} ($\Zo\Zo$, $\Wo\Zo$, $\Wo\Wo$, QCD production).
Signal events are generated using \POWHEG and a dedicated generator from ref.~\cite{Gao:2010qx}.
Parton distribution functions (PDF) are modeled using the parametrization
CTEQ6~\cite{CTEQ6:2004} at leading order (LO) and
CT10~\cite{Lai:2010vv} at next-to-leading order (NLO).
For both signal and background MC, events are simulated using
a \GEANTfour~\cite{GEANT4} based model of the CMS detector and processed using the
same reconstruction algorithms as used for data.

Muons are measured with the tracker and the muon system.
Electrons are detected as tracks in the tracker pointing to energy clusters in the ECAL.
Both muons and electrons are required to have a momentum transverse to the pp beam direction, \PT,
greater than 20\GeV and 10\GeV,
for the leading and subleading \PT lepton, respectively.
These requirements are tightened to 40\GeV and 20\GeV in the
analysis of the H candidates at higher masses.
Leptons are measured in the pseudorapidity range $|\eta|<2.4$ for muons,
and $|\eta|<2.5$ for electrons, although for electrons
the transition range between the barrel and endcap, $1.44<|\eta|<1.57$, is excluded.
Both the $\PT$ and $\eta$ requirements are consistent with those in the online
trigger selection requiring two charged leptons, either electrons or muons.
In the high-mass analysis, we also accept events selected with a single-muon trigger.
The details of electron and muon identification criteria are described
elsewhere~\cite{VBTF}.
Muons are required to be isolated from hadronic activity in the detector
by restricting the sum of transverse momentum or energy in the tracker, ECAL, and HCAL,
within a surrounding cone of $\Delta R \equiv \sqrt{(\Delta\eta)^2+(\Delta\phi)^2} <0.3$,
to be less than 15\% of the measured $\PT$ of the muon,
where $\Delta\eta$ and $\Delta\phi$ are the differences in pseudorapidity and
in azimuthal angle measured from the trajectory of the muon.
Electron isolation requirements are similar but vary depending on the shape of the electron shower.
In both cases the energy associated with the lepton is excluded from the isolation sum.

Jets are reconstructed with the particle-flow (PF) algorithm~\cite{CMS-PAS-PFT-09-001},
which is an event reconstruction technique with the aim of reconstructing all particles
produced in a given collision event through the combination of information
from all sub-detectors. Reconstructed particle candidates are clustered to form PF
jets with the anti-$k_T$ algorithm~\cite{Cacciari:2008gp, FastJet}
with the distance parameter $R=0.5$.
The HCAL, ECAL, and tracker data are combined in the PF algorithm to measure jets.
Jets that overlap with isolated leptons  within  $\Delta R=0.5$
are removed from consideration.

Jets are required to be inside the tracker acceptance, thus allowing high reconstruction efficiency
and precise energy measurements using PF algorithm.
Jet-energy corrections are applied to account for the non-linear response
of the calorimeters to the particle energies and other instrumental effects.
These corrections are based on in-situ measurements using dijet and $\Pgg+\text{jet}$ data
samples~\cite{Chatrchyan:2011ds}.
Overlapping minimum bias events (pile-up) coming from different proton-proton collisions
and the underlying event have an effect on jet reconstruction by contributing additional
energy to the reconstructed jets.
The median energy density resulting from pile-up is evaluated in each event,
and the corresponding energy is subtracted from each jet~\cite{Cacciari:2008gn}.
A jet requirement, primarily based on the energy balance between charged
and neutral hadrons in a jet, is applied to remove misidentified jets.
All jets are required to have $\PT>30\GeV$.

\begin{figure}[htbp]
\begin{center}
\centerline{
{\includegraphics[width=0.43\linewidth]{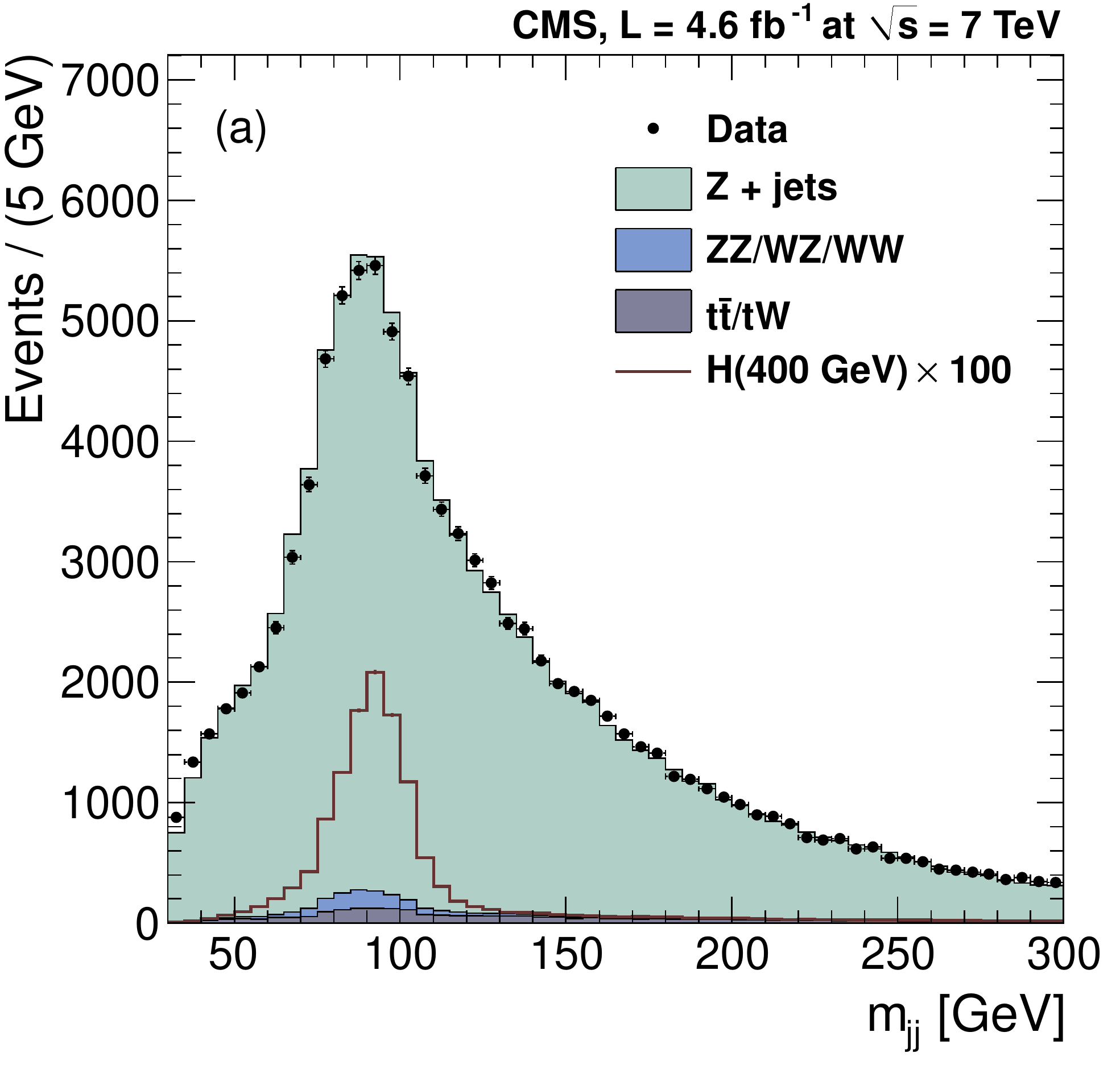}}
{\includegraphics[width=0.43\linewidth]{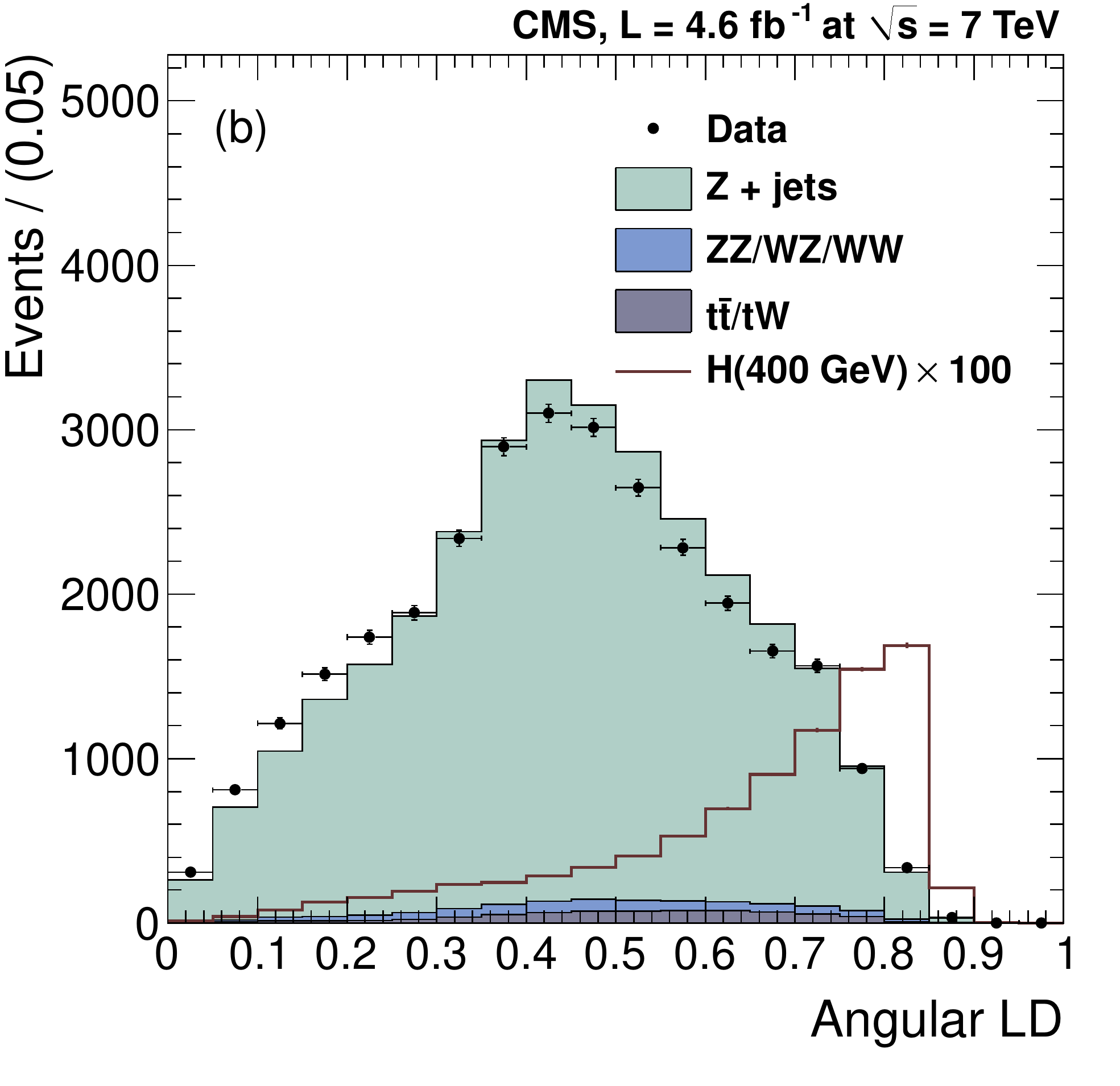}}
}
\centerline{
{\includegraphics[width=0.43\linewidth]{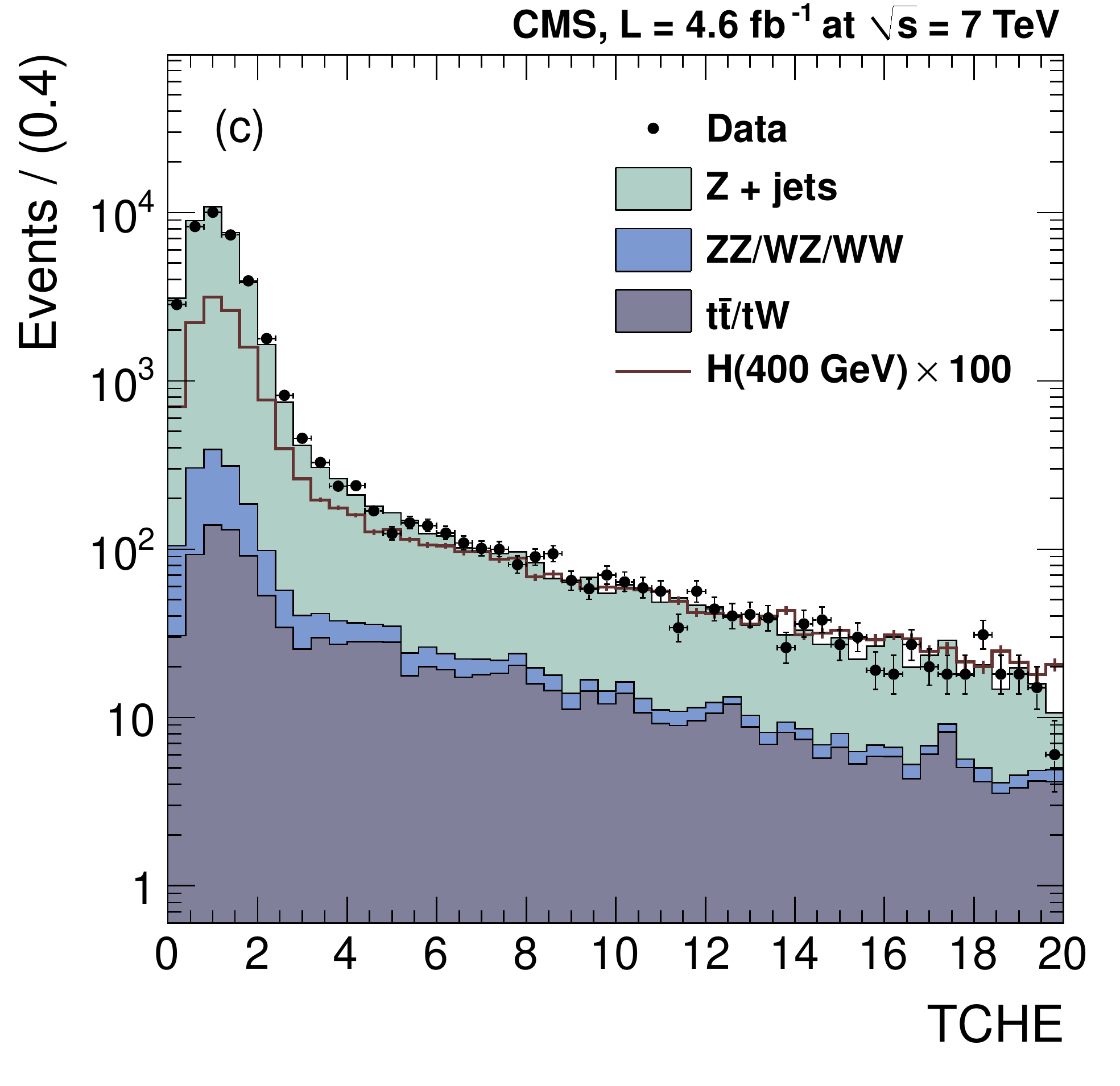}}
{\includegraphics[width=0.43\linewidth]{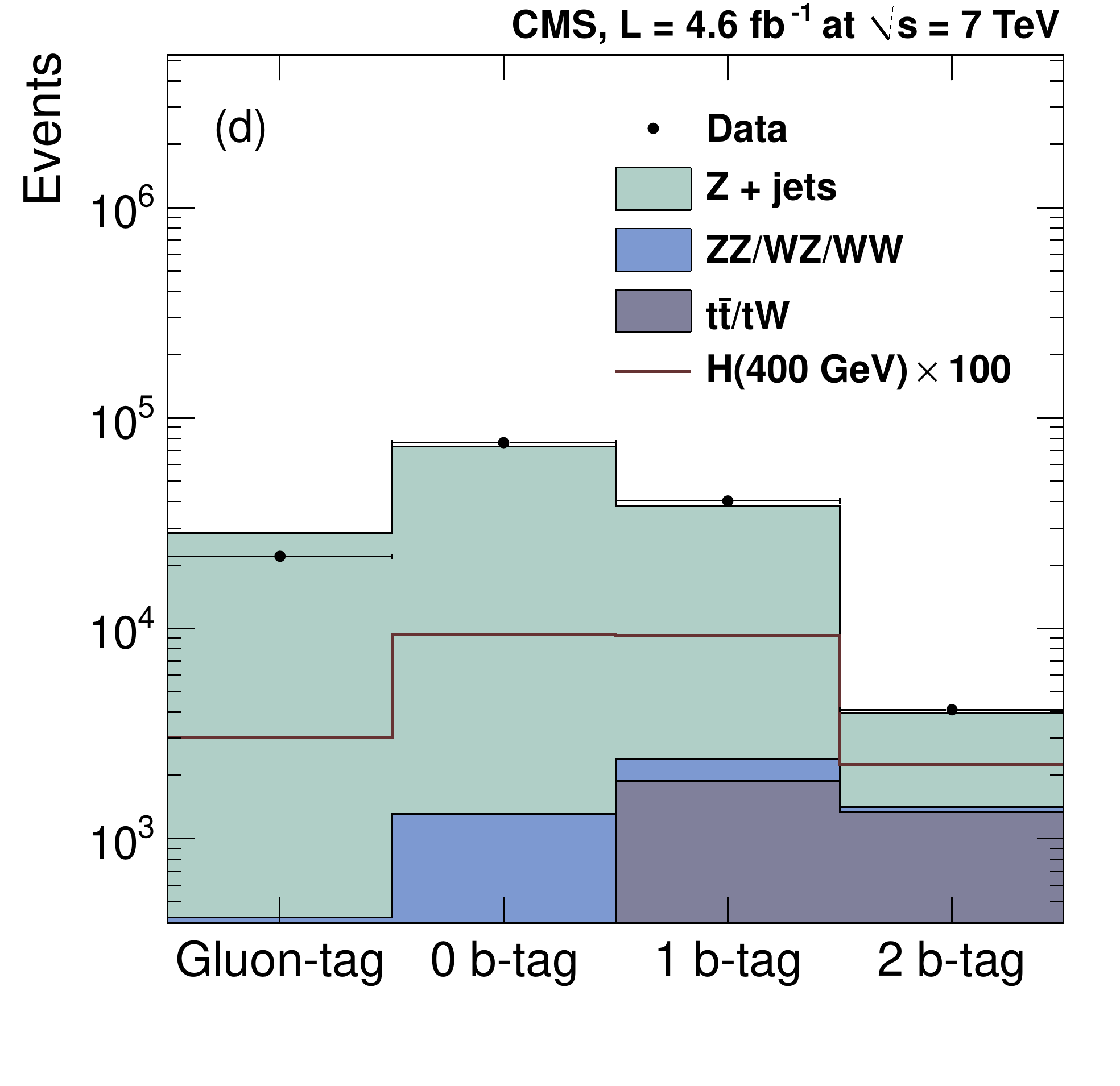}}
}
\centerline{
{\includegraphics[width=0.43\linewidth]{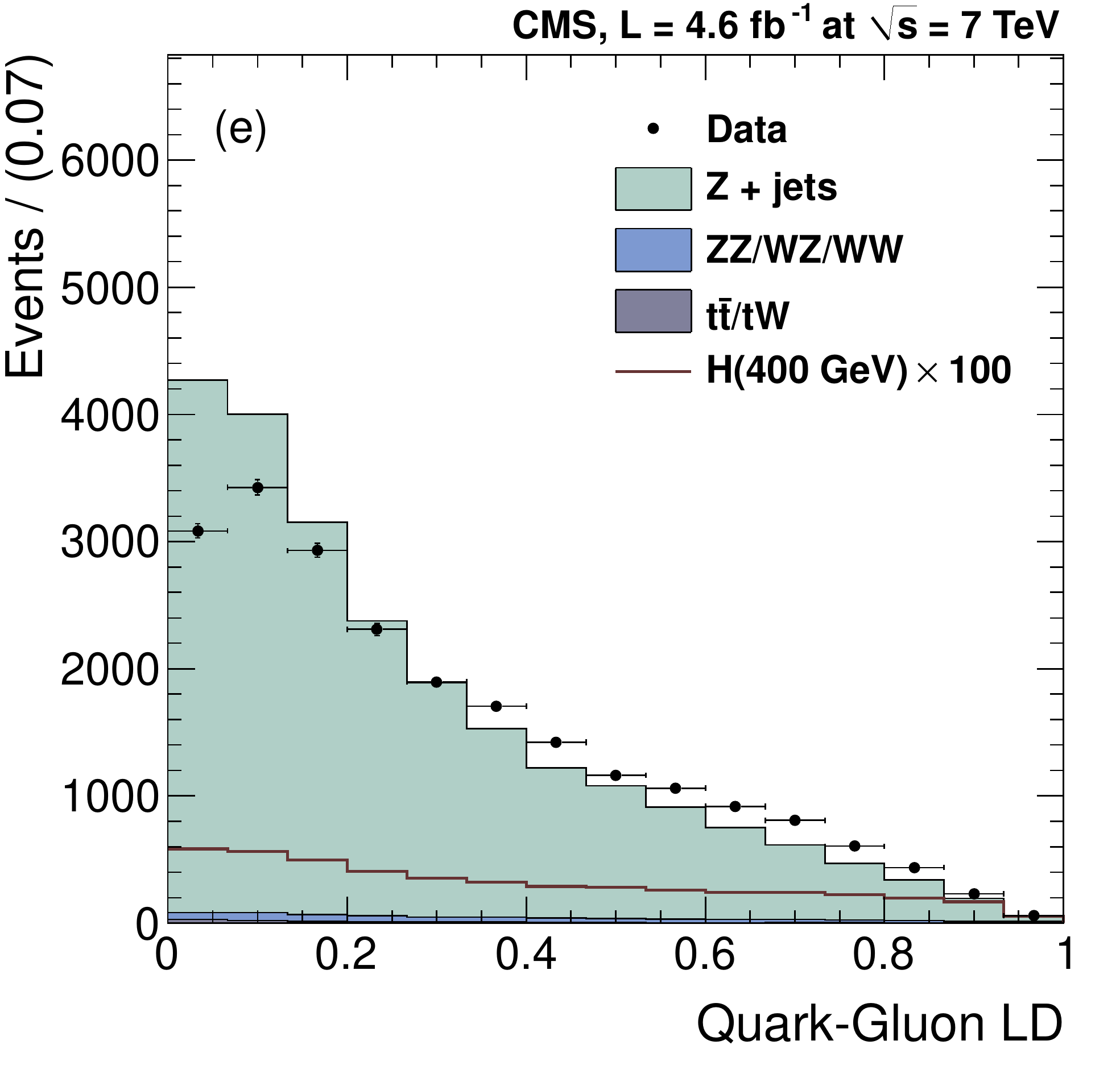}}
{\includegraphics[width=0.43\linewidth]{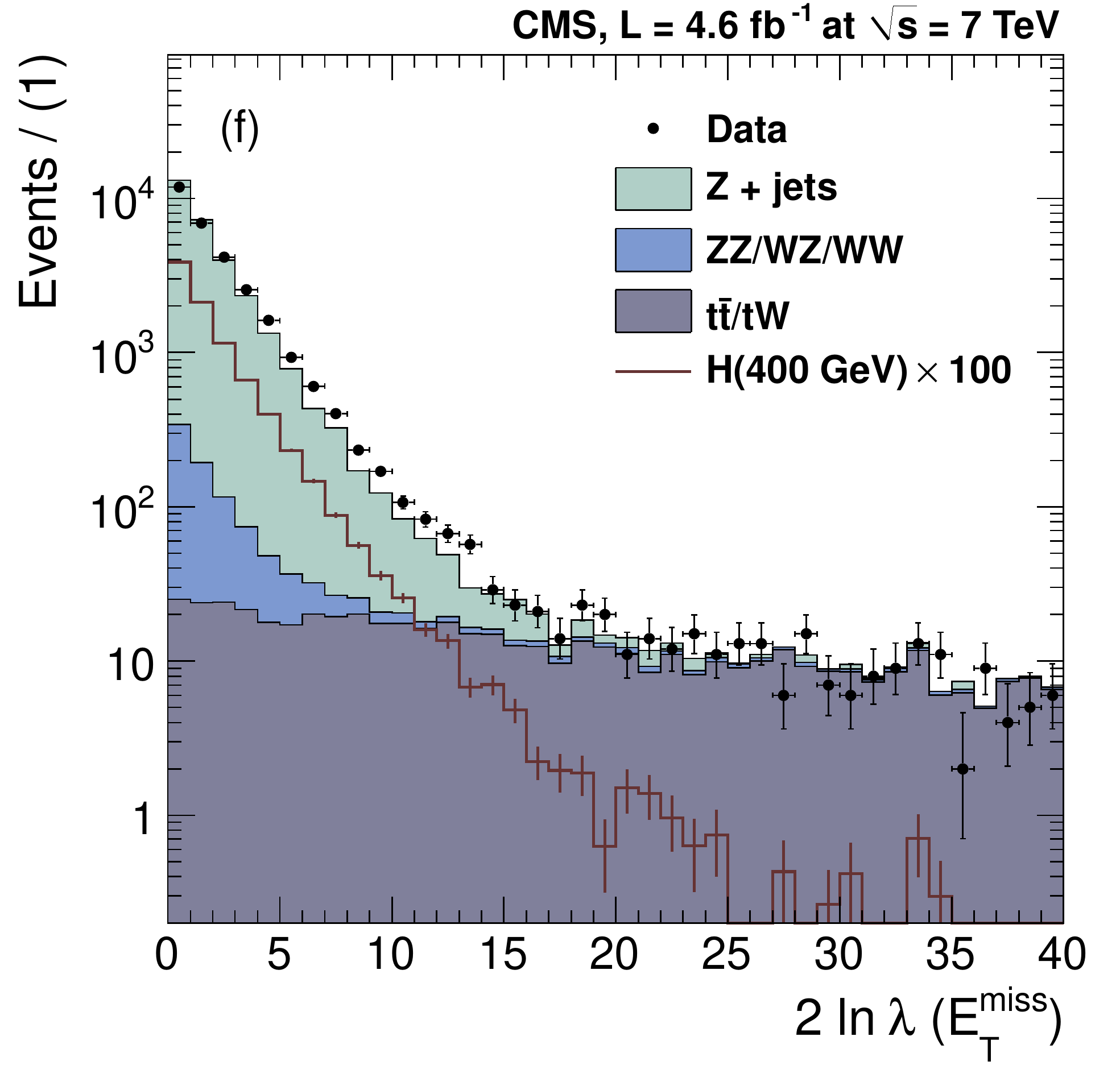}}
}
\caption{
Distribution of the dijet invariant mass $\mjj$ (a),
angular likelihood discriminant (b),
b-tagging discriminant (c),
flavor tagging category (d), including the gluon-tagged category,
quark-gluon likelihood discriminant (e), and $2\ln{\lambda} (\MET)$ (f).
Points with error bars show distributions of data after preselection
requirements defined in table~\ref{table-selection}
with an additional requirement $70<{ \mll}<110\GeV$.
Solid histograms depict the background expectation from simulated events
with the different components illustrated. Open histograms indicate the expected distribution
for a Higgs boson with a 400\GeV mass, multiplied by a factor of 100.
}
\label{fig:data_loose}
\end{center}
\end{figure}

Each pair of oppositely charged leptons and each pair of jets are considered as $\Zo$ candidates.
Background suppression is primarily based on the dilepton and dijet invariant masses,
$\mll$ and $\mjj$. The requirement $75<\mjj<105\GeV$
is applied in order to reduce the $\Zo$+jets background
and $70<\mll<110\GeV$ to reduce background without a $\Zo$ in the final state, such as $\TT$.
Figure~\ref{fig:data_loose}\,(a) shows the dijet invariant mass $\mjj$ distribution for signal and background.
In the search for the Higgs boson in the final state $\Zo\Zo^*$, we require
the invariant mass of the $\Zo^*\to\ell^-\ell^+$ candidate to be less than 80\GeV
instead of the previous requirement.
Below threshold for on-shell production of $\Zo\Zo$, the signal cross section is much smaller but also the
$\Zo^*/\Pgg^*$+jets background is strongly reduced.

The statistical analysis is based on the invariant mass of the Higgs boson candidate,
$\mZZ$, which is calculated using a fit of the final state four momenta and
applying the constraint that the dijet invariant mass is consistent with the mass of the $\Zo$ boson.
The experimental resolutions are taken into account in this fit.

Since the Higgs boson is spinless, the angular distribution of its decay products
is independent of the production mechanism.
Five angles $(\theta^*, \Phi_1, \theta_1, \theta_2, \Phi)$ defined in ref.~\cite{Gao:2010qx}
and in figure~\ref{fig:angles-HZZ2l2q} fully describe the kinematics
of the $\Pg\Pg\to\HZZllqq$ process.
Further kinematic selection exploits these five angular observables,
which are only weakly correlated with
the invariant masses of the H and the two $\Zo$ bosons
and with the longitudinal and transverse momenta of the Higgs boson candidate.
The five angles along with the invariant masses provide most of the discriminating power
between signal and background.
We construct an angular likelihood discriminant (LD) based on the probability ratio of the
signal and background hypotheses ${\cal P_\mathrm{sig}}/({\cal P_\mathrm{sig}}+{\cal P_\mathrm{sig}})$,
as described in ref.~\cite{Gao:2010qx}. The likelihood ratio is defined for each value of $\mZZ$
and its dependence on $\mZZ$ is parameterized with smooth functions.
Distributions of the angular LD for signal and background are shown in figure~\ref{fig:data_loose}\,(b).
The signal probability distribution is a correlated five-dimensional angular parameterization
multiplied by empirically determined polynomial acceptance functions from simulation
that describe non-uniform reconstruction efficiencies in the detector.
The background distribution is an empirical parameterization taken as a product
of independent distributions for each observable using simulation.
Both are parameterized as functions of $\mZZ$.
Cuts on the angular LD are chosen to optimize
the expected sensitivity to the production of a SM Higgs boson and depend on $\mZZ$.
The angular LD was found to have marginal separation power for $\mZZ<170\GeV$
and therefore is not used in selection requirements for this low-mass range.

The parton type of the jets provides a powerful tool for background discrimination.
In signal events, the jets originate from $\Zo$ bosons decaying to quarks that
subsequently hadronize. The flavor of quarks in $\Zo$ decays is almost equally
distributed among the five types \cPqd, \cPqu, \cPqs, \cPqc, \cPqb,
with some preference given to the down-type quarks.
The dominant background
is a leptonically decaying $\Zo$ boson produced in association with high-$\PT$ jets,
a process in which gluon radiation plays a major role. Beside gluons, the
\cPqu and \cPqd quarks from the protons dominate the jet production associated with the $\Zo$.
Therefore, the main features that discriminate signal from background are the relatively large
contribution of heavy-flavor quarks (\cPqb and \cPqs) and the absence of gluons. We take advantage
of both features in the analysis by tagging the b flavor and introducing a
likelihood discriminant that separates gluon and light-quark jets on a statistical basis,
as described below.

To identify jets originating from the hadronization of bottom quarks, we use the CMS track
counting high-efficiency (TCHE) \cPqb-tagging algorithm~\cite{CMS-BTV-10-001, CMS-BTV-11-003},
which relies on tracks with large impact parameters. A jet is \cPqb-tagged if
there are at least two tracks each with a three-dimensional
impact-parameter significance larger than a given threshold which has been optimized.
The distributions of the resulting \cPqb-tagging discriminant is shown in figure\,\ref{fig:data_loose}~(c).
The data are split into three \cPqb-tag categories:
a 2 \cPqb-tag category is required to have one jet identified with medium ($\sim$65\% efficiency)
and the other jet with loose ($\sim$80\% efficiency) TCHE requirements;
events not selected in the 2 \cPqb-tag category are categorized as 1 \cPqb-tag if they have at least
one jet satisfying the loose-tag requirements;
the 0 b-tag category contains all the remaining events.
The composition of the expected signal and background events varies significantly
among the three categories, see figure~\ref{fig:data_loose}\,(d).

The 0 \cPqb-tag category is dominated by the $\Zo$+jets background, and from these events we further
select a ``gluon-tagged" category, which is excluded from further analysis if the two leading
jets are consistent with being initiated by gluons, based on three measured quantities.
These are the number of charged hadronic particle tracks,
the number of photon and neutral hadrons, and the variable
${\rm PTD}=\sqrt{ {\sum p_{\mathrm{\mathrm{T}}}^2}/{\left ( \sum p_{\mathrm{\mathrm{T}}} \right )^2} }$,
where the sum is extended over all jet constituents.
The variable PTD is related to the fragmentation variable
$z = p_{\mathrm{\mathrm{T}}}(\text{constituent})/p_{\mathrm{\mathrm{T}}}(\text{jet})$
and is approximately equal to $\sqrt{\sum z^2}$.
Gluon hadronization favors the production of a larger number of stable particles.
This translates into the observation of softer (low PTD),
high-multiplicity jets when compared to those generated by final-state quarks.
We construct a quark-gluon LD from the above three observables.
The corresponding LD distributions for signal and background are shown in
figure~\ref{fig:data_loose}\,(e).
The relative number of gluon- and quark-jets for the main background,
$\Zo$+jets, is not well known and it is not expected to be well reproduced by the simulation.
The quark-gluon LD is instead verified using data samples of \Pgg+jets enriched in quark-jets.

In order to suppress the substantial $\ttbar$ background in the 2 \cPqb-tag category,
we apply a selection on the missing transverse energy ($\MET$) which is defined as
the modulus of the negative vector sum of all reconstructed PF particles in the event.
We construct a discriminant, $\lambda$, which is the ratio of the likelihoods of the hypothesis
with $\MET$ equal to the value measured with the PF algorithm and the null
hypothesis ($\MET=0$)~\cite{Chatrchyan:2011tn}.
This discriminant provides a measure that the event contains genuine missing transverse energy.
The distribution of $2\ln{\lambda} (\MET)$ is shown in figure~\ref{fig:data_loose}\,(f).
We apply a loose requirement, $2\ln{\lambda} (\MET) < 10$, in the 2 \cPqb-tag category only.
In the low-mass analysis, we instead apply the selection requirement $\MET < 50\GeV$
in the 2 \cPqb-tag category.

Data and MC predictions of background distributions after the preselection
requirements are shown in figure~\ref{fig:data_loose}, where the additional contribution
of a Higgs boson signal would be indistinguishable above the overwhelming background.
The overall agreement between background simulation and data is good
except for systematic differences related to the quark-gluon composition
in $\Zo$+jets events, as shown in figure~\ref{fig:data_loose}\,(e).
We do not rely directly on simulation for background estimates.
Instead, the background is determined directly using sidebands in data (see Section~\ref{sec:analysis}).

The main selection requirements are summarized in table~\ref{table-selection}.
When an event contains multiple candidates passing the selection requirements,
we retain the one with jets in the highest \cPqb-tag category for the analysis.
Further ambiguity between multiple candidates is resolved selecting the candidate
with $\mjj$ and $\mll$ values closest to the $\Zo$ boson mass \mZ{}.
The distribution of the $\mZZ$ invariant mass for background and
data are displayed for the three \cPqb-tag  categories
in figure~\ref{fig:mZZ_kinfit_hiMass}. No significant deviation is observed between the data
and the expectation for background. The main backgrounds include
inclusive $\Zo$ production with either light-flavor or heavy-flavor jets, top-quark production,
and diboson production such as $\Wo\Zo$ and $\Zo\Zo$.
The expected and observed event yields are listed in table~\ref{table-yields}.
The expected background is quoted from the $\mjj$ sideband procedure described below and from simulation.
In the low-mass range, the background distribution is obtained from the $\mjj$ sideband while
its size is estimated from the $\mZZ$ sideband chosen for each $\mH$ hypothesis, as discussed below.

\section{Event Analysis}
\label{sec:analysis}
\begin{figure*}[htbp]
\begin{center}
\centerline{
{\includegraphics[width=0.43\textwidth]{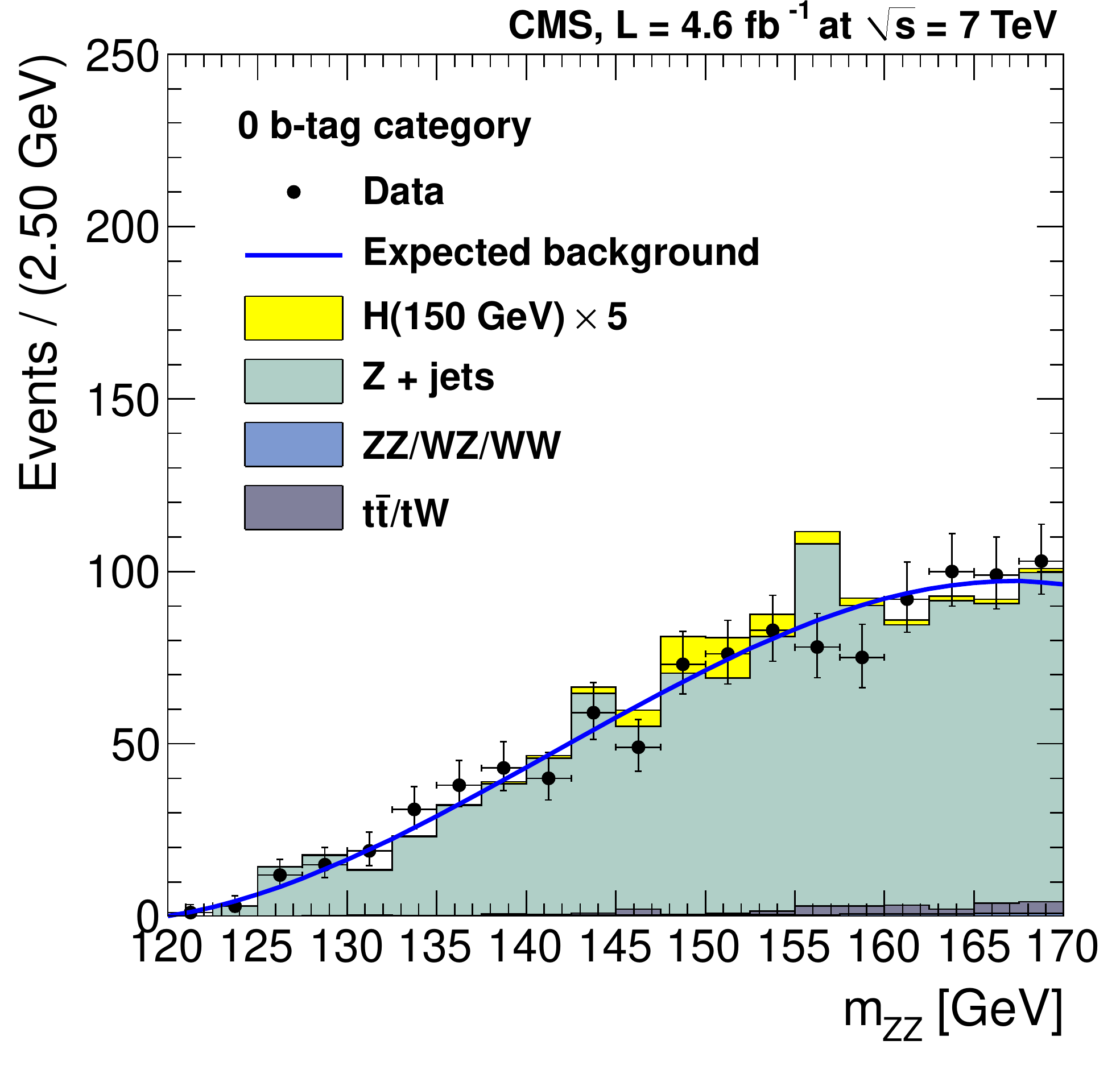}}
{\includegraphics[width=0.43\textwidth]{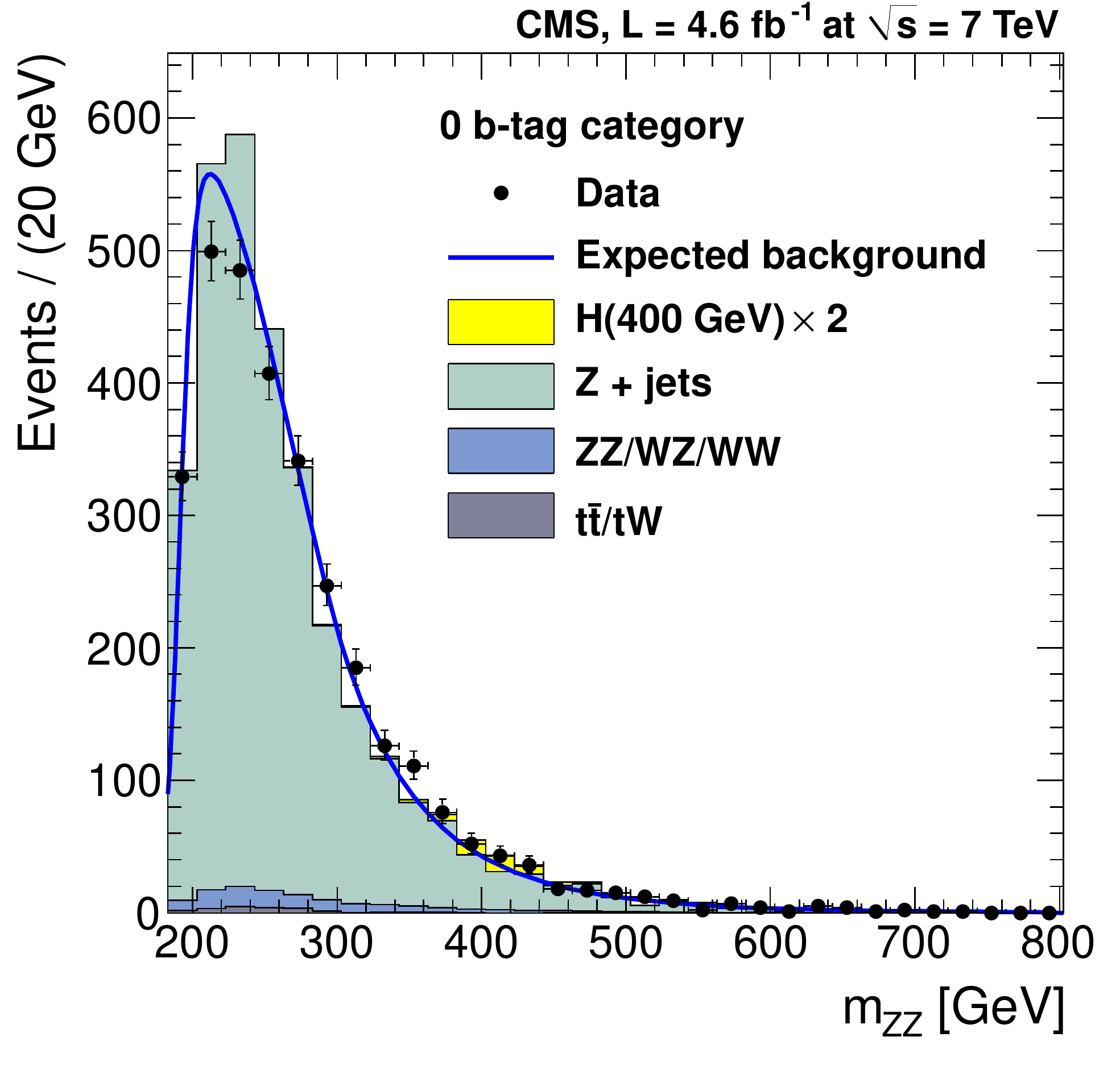}}
}
\centerline{
{\includegraphics[width=0.43\textwidth]{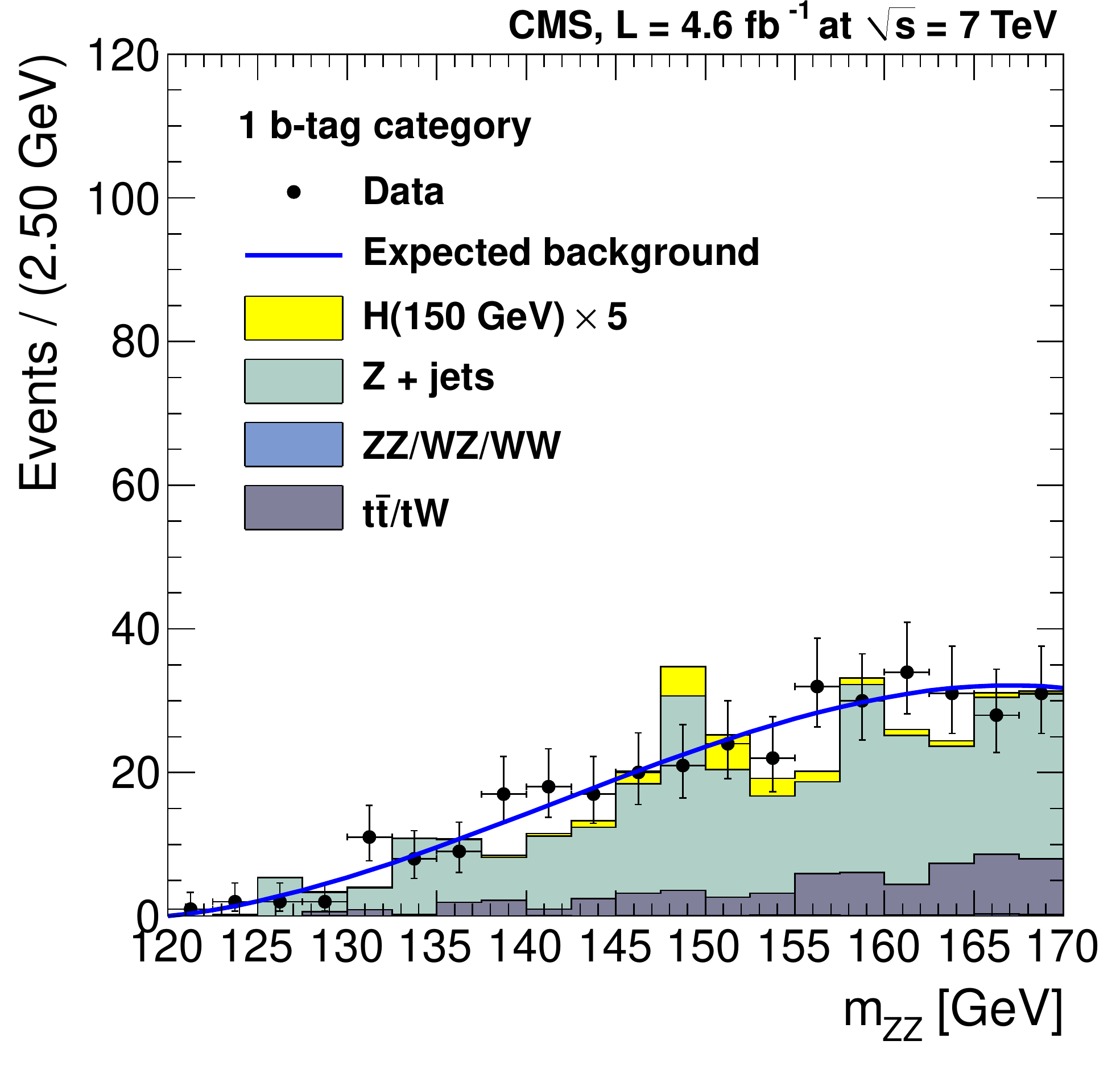}}
{\includegraphics[width=0.43\textwidth]{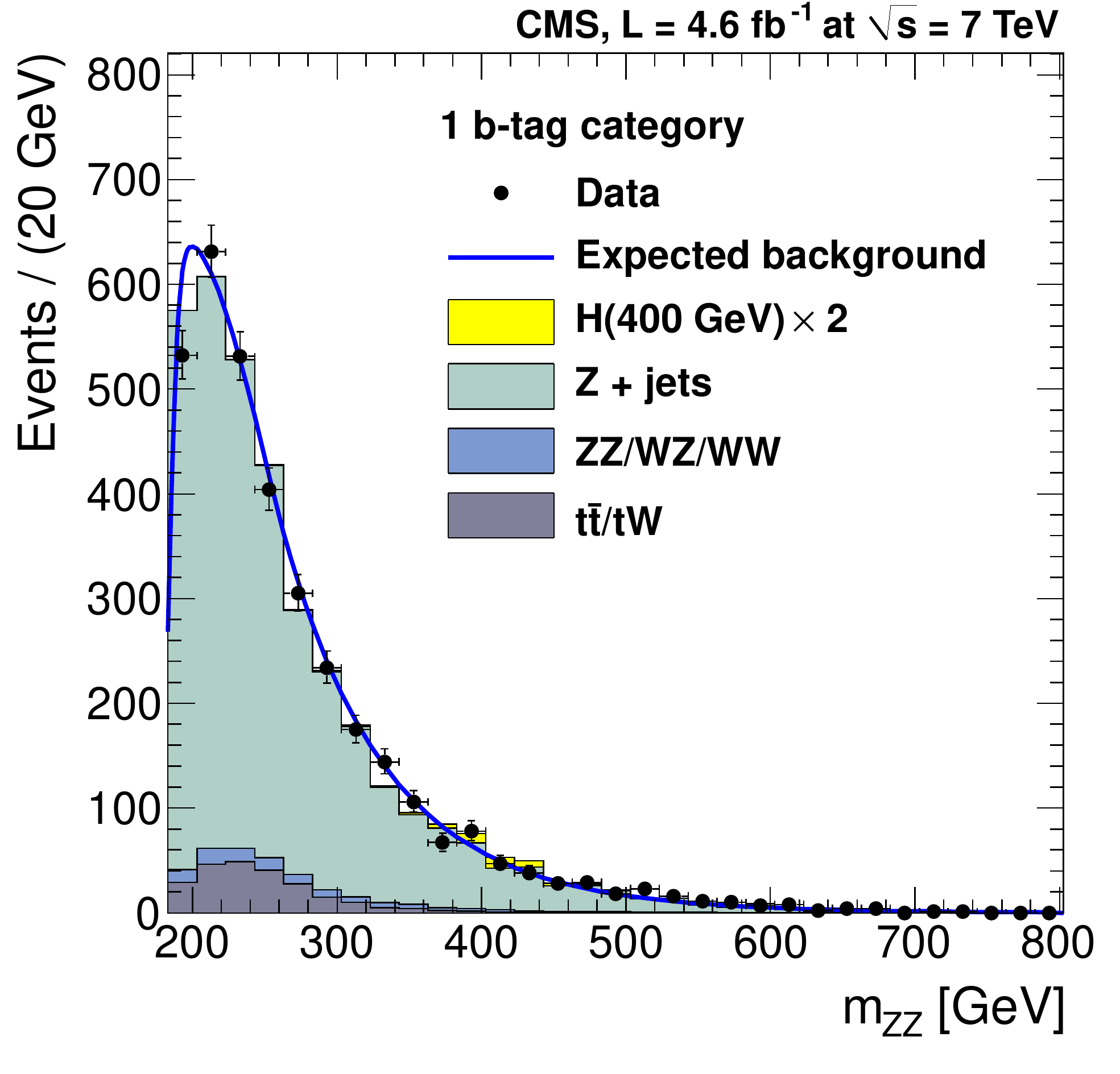}}
}
\centerline{
{\includegraphics[width=0.43\textwidth]{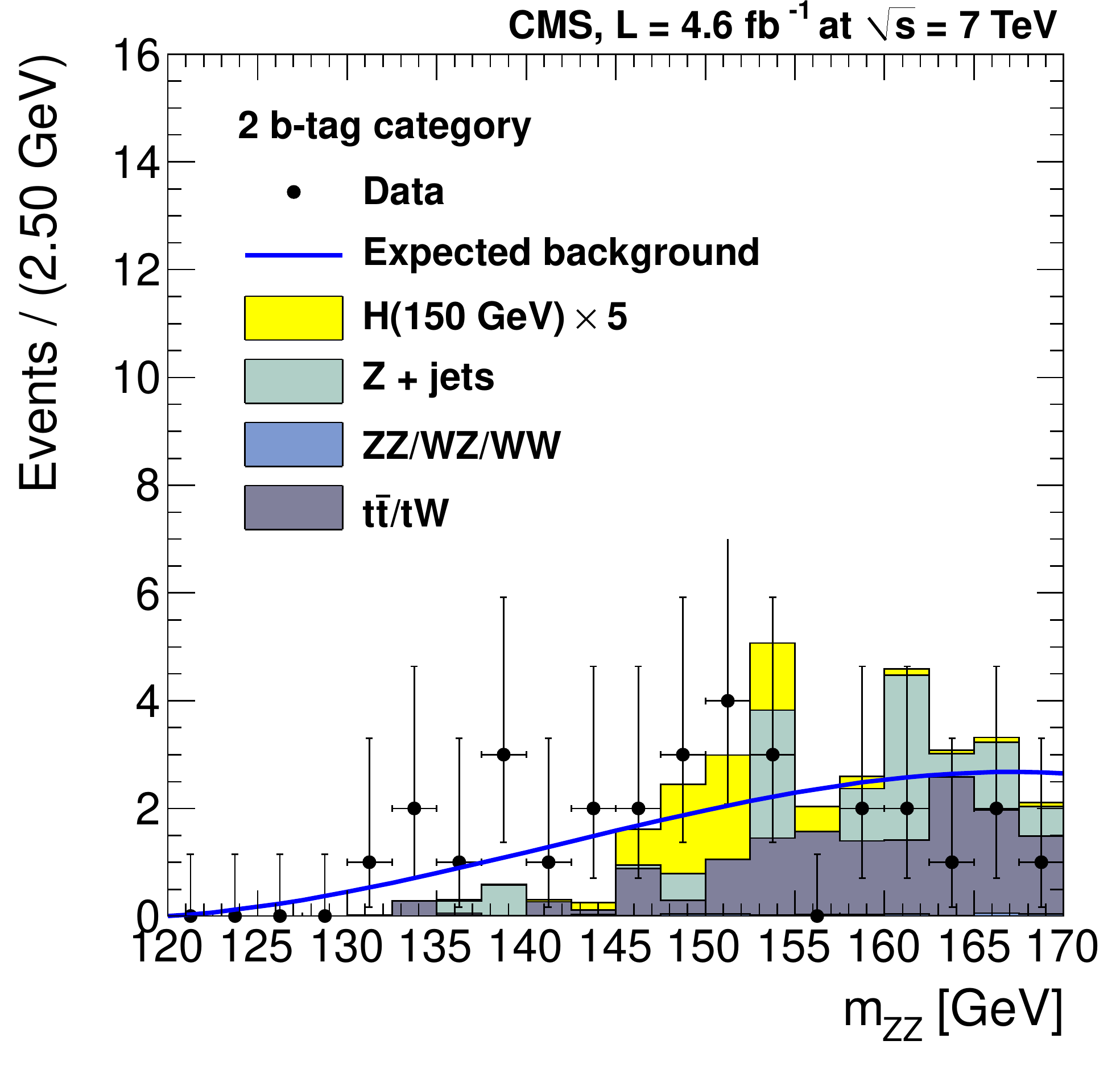}}
{\includegraphics[width=0.43\textwidth]{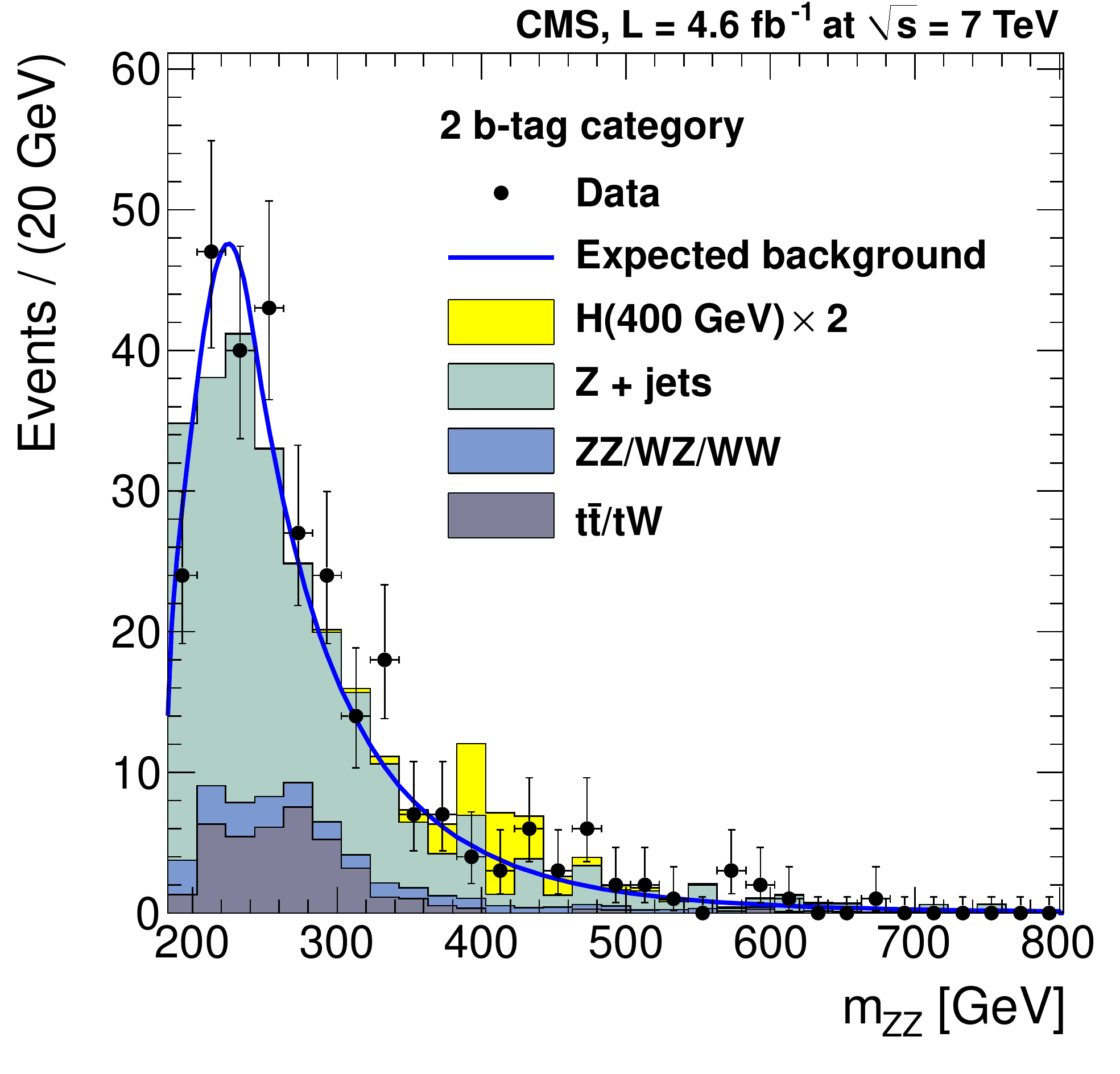}}
}
\caption{
The $\mZZ$ invariant mass distribution after final selection in three categories:
0~\cPqb-tag~(top),
1~\cPqb-tag~(middle), and
2~\cPqb-tag~(bottom).
The low-mass range $120<\mZZ<170\GeV$ is shown on the left
and the high-mass range $183<\mZZ<800\GeV$ is shown on the right.
Points with error bars show distributions of data and
solid curved lines show the prediction of background from the sideband extrapolation procedure.
In the low-mass range, the background is estimated from the $\mZZ$ sideband for each Higgs
mass hypothesis and the average expectation is shown.
Solid histograms depicting the background expectation from
simulated events for the different components are shown.
Also shown is the SM Higgs boson signal with the mass of 150 (400)\GeV and cross section
5 (2) times that of the SM Higgs boson, which roughly corresponds to expected exclusion
limits in each category.
}
\label{fig:mZZ_kinfit_hiMass}
\end{center}
\end{figure*}

Data containing a Higgs boson signal would have a distinct resonance peak in addition
to the continuum background distribution.
The estimates from simulation shown in figure~\ref{fig:mZZ_kinfit_hiMass}
provide a good illustration of the expected background but require further validation
of both theoretical predictions, such as production cross section,
and detector effects, e.g. \cPqb-tagging efficiency.
These effects can explain the discrepancies between data and background simulation,
which are sizable near the $\Zo\Zo$ threshold around $\mZZ=200\GeV$.
However, the analysis technique relies on sidebands measured in
data and is largely insensitive to the modeling of the $\mZZ$ distributions.

\begin{table}[htbp]
\begin{center}
\caption{Summary of kinematic and topological selection requirements.
Numbers in parentheses indicate additional selection requirements in the $\mZZ$ range
[125, 170]\GeV, where angular and quark-gluon likelihood discriminant requirements are not used.
}
\label{table-selection}
\vspace*{\medskipamount}
\begin{tabular}{|l|c|c|c|}
\cline{2-4}
 \multicolumn{1}{c|}{} & \multicolumn{3}{c|}{ preselection }\\
\hline
${\PT(\ell^\pm)}$  & \multicolumn{3}{c|}{leading $\PT > 40(20)\GeV$, subleading $\PT > 20(10)\GeV$}\\
${\PT(\rm jets)}$ & \multicolumn{3}{c|}{$>30\GeV$} \\
${ |\eta|(\ell^\pm)}$ &  \multicolumn{3}{c|}{$<2.5~(\Pe^\pm)$, $<2.4~(\mu^\pm)$} \\
${ |\eta|(\text{jets})}$ &  \multicolumn{3}{c|}{$<2.4$} \\
\hline
 \multicolumn{1}{c}{\vspace{-0.2cm}} & \multicolumn{3}{c}{\vspace{-0.2cm}}\\
\cline{2-4}
 \multicolumn{1}{c|}{} & \multicolumn{3}{c|}{ final selection }\\
\cline{2-4}
\multicolumn{1}{c|}{} & 0 \cPqb-tag & 1 \cPqb-tag  & 2 \cPqb-tag \\
\hline\hline
\cPqb-tag & none & one loose &  medium\,\&\,loose\\
angular LD   & $> 0.55+ 0.00025\,\mZZ$ & $>0.302+0.000656\,\mZZ$ & $> 0.5$ \\
quark-gluon LD &  $>0.10$ & none & none  \\
$\MET$ requirements  &  none & none &  $2\ln{\lambda} (\MET)<10$ \\
  &  &  &   ($\MET<50\GeV$) \\
\hline
${ \mjj}$  &  \multicolumn{3}{c|}{$\in$ [75, 105]\GeVnn} \\
${ \mll}$   & \multicolumn{3}{c|}{$\in$ [70, 110] ($<$80)\GeV} \\
${ \mZZ}$  &  \multicolumn{3}{c|}{$\in$  [183, 800] ($\in$ [125, 170])\GeV}\\
\hline
\end{tabular}
\end{center}
\end{table}
\begin{table}[htbp]
\begin{center}
\caption{
Observed and expected event yields for 4.6\fbinv of data.
The yields are quoted in the range $125<\mZZ<170\GeV$
or $183<\mZZ<800\GeV$, depending on the Higgs boson mass hypothesis.
The expected background is quoted from the $\mjj$ sideband procedure and from simulation (MC).
In the low-mass range, the background is estimated from the
$\mZZ$ sideband for each Higgs mass hypothesis and is not quoted in the table.
The errors on the expected background from simulation include only statistical uncertainties.
}
\label{table-yields}
\vspace*{\medskipamount}
\begin{tabular}{|c|c|c|c|}
\cline{2-4}
 \multicolumn{1}{c|}{}  & 0 \cPqb-tag & 1 \cPqb-tag  & 2 \cPqb-tag \\
\hline\hline
 \multicolumn{4}{|c|}{  ${ \mZZ}\in$ [125, 170] }   \\
 \hline
 observed yield      & 1087 & 360 & 30  \\
 expected background ($\mjj$ sideband)   & $1050\pm 54$ & $324\pm 28$  &  $19\pm 5$  \\
 expected background (MC)  & $1089\pm 39$ & $313\pm20$  & $24\pm 4$  \\
\hline \hline
 \multicolumn{4}{|c|}{ ${ \mZZ}\in$ [183, 800]  }   \\
 \hline
 observed yield               & 3036         & 3454         & 285\\
 expected background ($\mjj$ sideband)   & $3041\pm54$  & $3470\pm59$  & $258\pm17$  \\
 expected background (MC)     & $3105\pm39$  & $3420\pm41$  & $255\pm11$  \\
\hline \hline
 \multicolumn{4}{|c|}{ signal expectation (MC) }   \\
 \hline
  \mH=150  \GeV        & 10.1 $\pm$ 1.5 & 4.1 $\pm$ 0.6 & 1.6 $\pm$ 0.3\\
  \mH=250  \GeV        & 24.5 $\pm$ 3.5  & 21.7 $\pm$ 3.0  & 8.1  $\pm$ 1.7 \\
  \mH=350  \GeV        & 29.6 $\pm$ 4.3  & 26.0 $\pm$ 3.7  & 11.8 $\pm$ 2.5 \\
  \mH=450  \GeV        & 16.5 $\pm$ 2.4  & 15.8 $\pm$ 2.2  & 7.9  $\pm$ 1.7 \\
  \mH=550  \GeV        &  6.5 $\pm$ 1.0  & 6.5  $\pm$ 0.9  & 3.6  $\pm$ 0.8 \\
\hline
\end{tabular}
\end{center}
\end{table}

In order to minimize the systematic uncertainty from the background models, we estimate the background distribution
from the  $\mjj$ sidebands, defined as  $60<\mjj<75\GeV$ and $105 <\mjj<130\GeV$.
In simulation, the composition and distribution of the dominant backgrounds in the sidebands
is similar to that in the signal region, $75<\mjj<105\GeV$.
The expected number of background events, $N_\text{bkg}(\mZZ)$, is
obtained from the number of events in the sidebands, $N_\mathrm{sb}(\mZZ)$, as follows:
\begin{eqnarray}
N_\text{bkg}(\mZZ)
=N_\mathrm{sb}(\mZZ)\times\frac{N^\text{sim}_\text{bkg}(\mZZ)}{N^\text{sim}_\mathrm{sb}(\mZZ)}
=N_\mathrm{sb}(\mZZ)\times\alpha(\mZZ),
\end{eqnarray}
where $\alpha(\mZZ)$ is the ratio of the expected number of background events
in the signal and sideband regions obtained from simulation. This factor corrects
for acceptance differences between the two regions and also for differences in
background composition.

In the high-mass range, the distributions derived from data sidebands are measured
for each of the three \cPqb-tag requirements and give the normalization of the background
and its dependence on $\mZZ$. The correction $\alpha(\mZZ)$ reaches a
maximum of about 1.2 near the threshold of 2$\mZ$ and falls to nearly a
constant value between 0.75 and 1.0 elsewhere, depending on \cPqb-tag
and kinematic requirements.

In the low-mass range, below the
2$\mZ$ threshold, the same kinematic selections are applied to all
\cPqb-tag categories and a single background spectrum is derived from the $\mjj$ sidebands.
The correction $\alpha(\mZZ)$ is not applied, and instead the normalizations in each category are obtained
as a function of $\mH$, using an $\mZZ$ sideband outside the window $\mH\pm5\GeV$.

The results of the sideband extrapolation procedures
are shown as solid curves in figure~\ref{fig:mZZ_kinfit_hiMass}
and are in good agreement with the observed distributions in data.
In all cases, the dominant backgrounds
include $\Zo$+jets with either light- or heavy-flavor jets and top background, both of which
populate the $\mjj$ signal region and the $\mjj$ sidebands.
The diboson background amounts to less than 5$\%$ of the total in the 0 and
1 \cPqb-tag categories and about 10\% in the 2 \cPqb-tag category.
This diboson background is accounted for by $\alpha(\mZZ)$ in the high-mass range
and by the $\mZZ$ sideband procedure in the low-mass range.

The distribution of $\mZZ$ for the background is parameterized with an empirical function,
fitted to the shape and normalization determined from the sidebands.
The advantage of this approach is that most
of the systematic uncertainties on the background cancel.
The dominant normalization uncertainty in the background estimation
is due to statistical fluctuations of the number of events in the sidebands.
The reconstructed signal distributions are described with a two-sided Crystal-Ball function~\cite{CrystalBall}
and an empirical function reflecting misreconstruction of the Higgs boson decay products.
The signal reconstruction efficiency and the $\mZZ$ distribution are parameterized as a function of $\mH$
and are extrapolated to all mass points. The main uncertainties in the signal $\mZZ$ parameterization
are due to resolution which is predominantly affected by the uncertainty on the jet energy scale~\cite{Chatrchyan:2011ds}.

The $\mZZ$ distributions of the selected events are split into six categories based
on the \cPqb-tag type and the lepton flavor. These events are examined for
43 hypothetical Higgs boson masses in a range between 130\GeV and 164\GeV,
and 73 hypothetical Higgs boson masses in the range between 200\GeV and 600\GeV,
where the mass steps are optimized to account for the expected width, $\Gamma_{\Ho}$, and
resolution for measurement of $\mH$~\cite{LHC-HCG}.
For each mass hypothesis, we perform a simultaneous likelihood fit of the six $\mZZ$ distributions
using the statistical approaches discussed in ref.~\cite{LHC-HCG}.
As an alternative, we have also studied a cut-based analysis
that counts events in regions of the $\mZZ$ distribution and found consistent, but
systematically higher median expected limits compared to the likelihood fit approach.
We adopt the modified frequentist construction $\mathrm{CL_s}$~\cite{Junk,LEP-CLs,LHC-HCG}
as the primary method for reporting limits. As a complementary method to the frequentist paradigm,
we use the Bayesian approach~\cite{Bayes} and find consistent results.

\begin{table}[t]
\caption{
Summary of systematic uncertainties on signal normalization. Most sources give multiplicative uncertainties
on the cross-section measurement, except for the expected Higgs boson production cross section,
which is relevant for the measurement of the ratio to the SM expectation. The ranges indicate
dependence on $\mH$.
}
\label{table-systematics}
\begin{center}
\small
\begin{tabular}{|l|c|c|c|}
\hline
 source      &   0 \cPqb-tag   &   1 \cPqb-tag  &   2 \cPqb-tag \\
\hline
\hline
muon  reconstruction    &  \multicolumn{3}{c|}{2.7\%}   \\
\hline
electron reconstruction &  \multicolumn{3}{c|}{4.5\%}     \\
\hline
jet reconstruction             &  \multicolumn{3}{c|}{1--8\%}\\
\hline
pile-up              &  \multicolumn{3}{c|}{3--4\%}    \\
\hline
$\MET$                 &  -- & -- & 3--4\%  \\
\hline
\cPqb-tagging    &  2--7\% & 3--5\% & 10--11\%  \\
\hline
gluon-tagging    & 4.6\%   & -- & -- \\
\hline
acceptance (HqT)&  2\% & 5\% & 3\%  \\
\hline
acceptance (PDF)&  \multicolumn{3}{c|}{3\%}  \\
\hline
acceptance (VBF)&   \multicolumn{3}{c|}{1\%}  \\
\hline
signal cross section (PDF) &  \multicolumn{3}{c|}{ 8--10$\%$ }    \\
\hline
signal cross section (scale) &  \multicolumn{3}{c|}{ 8--11$\%$ }    \\
\hline
signal shape     &  \multicolumn{3}{c|}{ $1.5\times10^{-7}\% \times \mH^3$ [\GeV{}] }    \\
\hline
luminosity            &  \multicolumn{3}{c|}{4.5$\%$}   \\
\hline
\end{tabular}
\end{center}
\end{table}

The systematic uncertainties on signal normalization are summarized in table~\ref{table-systematics}.
We consider effects from lepton energy scale, resolution, selection, and trigger (electron/muon reconstruction);
jet resolution and efficiency (jet reconstruction); pile-up; $\MET$ requirements; heavy-quark flavor tagging and quark-gluon discrimination;
Higgs boson production mechanism; cross section and branching fractions; resonance mass shape;
and LHC luminosity.
Reconstruction efficiencies for leptons and their uncertainties are evaluated from data with a ``tag-and-probe''~\cite{VBTF}
approach where one lepton from an inclusive sample of $\Zo$ decays serves as a tag and the efficiency
for the reconstruction of the other lepton is calculated.
Contributions from jet reconstruction are evaluated by variation of the jet energy and resolution
within calibration uncertainties.
The contributions from the uncertainty on pile-up are taken from
the simulated difference between the reconstruction and the selection efficiency
with pile-up below and above the average expected value,
distributed according to the measurement in data.
The uncertainty of the $\MET$ selection efficiency is computed by examining the $\MET$
distributions from $\Zo$ inclusive production in MC simulation and
in data after subtraction of background from top production.
Uncertainties due to \cPqb\ tagging have been evaluated with a sample of jet
events enriched in heavy flavor by requiring a muon to be spatially close to a jet.
The uncertainty on the quark-gluon LD selection efficiency was evaluated using the $\Pgg+\text{jet}$
sample in data, which predominantly contains quark jets.

Uncertainties in the production mechanism affect the signal acceptance in the detector.
Both the longitudinal momentum of the Higgs boson, because of PDFs,
and the transverse momentum of the Higgs boson, because of QCD initial-state radiation effects,
are model dependent.
We rescale the transverse momentum distribution of the Higgs boson using the {HqT}~\cite{Bozzi:2005wk}
code and take the full change in the efficiency as a systematic uncertainty.
We follow the PDF4LHC~\cite{Botje:2011sn,Alekhin:2011sk,Lai:2010vv,Martin:2009iq,Ball:2011mu}
recommendation to estimate the uncertainty due to PDF knowledge and to calculate
the uncertainty on signal acceptance.
Uncertainties on the production cross section for the Higgs boson are taken from
ref.~\cite{LHCHiggsCrossSectionWorkingGroup:2011ti},
which includes uncertainties from the QCD renormalization and factorization scales, PDFs, and $\alpha_s$.
These uncertainties are separated between the gg and VBF production
mechanisms, but uncertainties on the \cPg\cPg\ process dominate in the total production cross section.
We also account for a small uncertainty because of a difference in signal acceptance with the gg and VBF
production mechanisms, while the selection efficiency was optimized and evaluated for the dominant gg production.
A relative uncertainty of 4.5\% on luminosity is applied to the signal normalization.

Recent studies \cite{Passarino:2010qk,LHCHiggsCrossSectionWorkingGroup:2011ti,Anastasiou:2011pi}
show that current Monte Carlo simulations do not describe the correct Higgs boson mass line shape
above $\approx 300\GeV$.
These effects are estimated to lead to an additional uncertainty on the theoretical cross section
of 10--30\% for $m_{\PH}$ of 400--600\GeV and are included in the calculations of the limits.

We also consider the production and decay of the Higgs boson within a model with four generations
of fermions (SM4)~\cite{Schmidt:2009kk,Li:2010fu,Anastasiou:2011pi, Denner:2011vt, Passarino:2011kv},
including electroweak radiative corrections.
The following scenario has been adopted in the SM4 calculations:
 $m_{\cPqb^\prime} = 600\GeV$ and
 $m_{\cPqt^\prime} - m_{\cPqb^\prime} = 50(1+0.2\ln(\mH/115))\GeV$,
following recommendation of ref.~\cite{LHCHiggsCrossSectionWorkingGroup:2011ti}.
The main difference from the SM is a higher production rate and somewhat different
branching fractions of the SM4 Higgs boson.
We assume that the main uncertainties on the SM4 Higgs production cross section
are the same as the gluon-fusion mechanism in the SM~\cite{LHCHiggsCrossSectionWorkingGroup:2011ti}.

In order to infer the presence or absence of a signal in the data sample,
we construct an appropriate test statistic $q$, a single number encompassing information
on the observed data, expected signal, expected background, and all uncertainties associated
with these expectations~\cite{LHC-HCG}.
The definition of $q$ makes use of a likelihood ratio for the signal+background model
and the model with the best-fit signal strength plus background.
We compare the observed value of the test statistic with its distributions expected under
the background-only and signal+background hypotheses.
The expected distributions are obtained by generating pseudo-datasets.
The signal strength which leads to a 95\% CL limit is determined for each Higgs mass
hypothesis under study.

\section{Results}
\label{sec:results}

\begin{figure}[htbp]
\begin{center}
\centerline{
\includegraphics[width=0.5\linewidth]{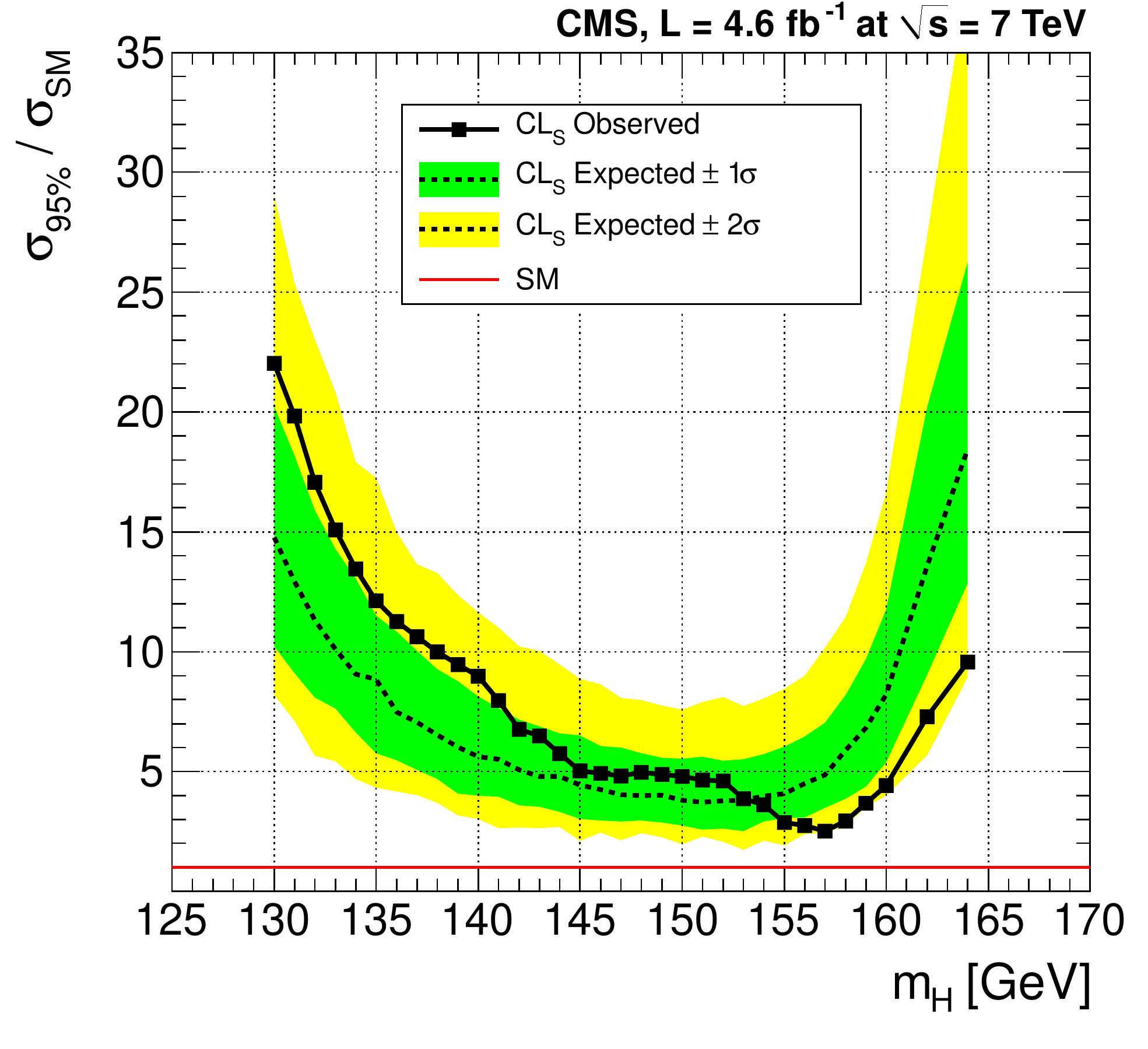}
\includegraphics[width=0.5\linewidth]{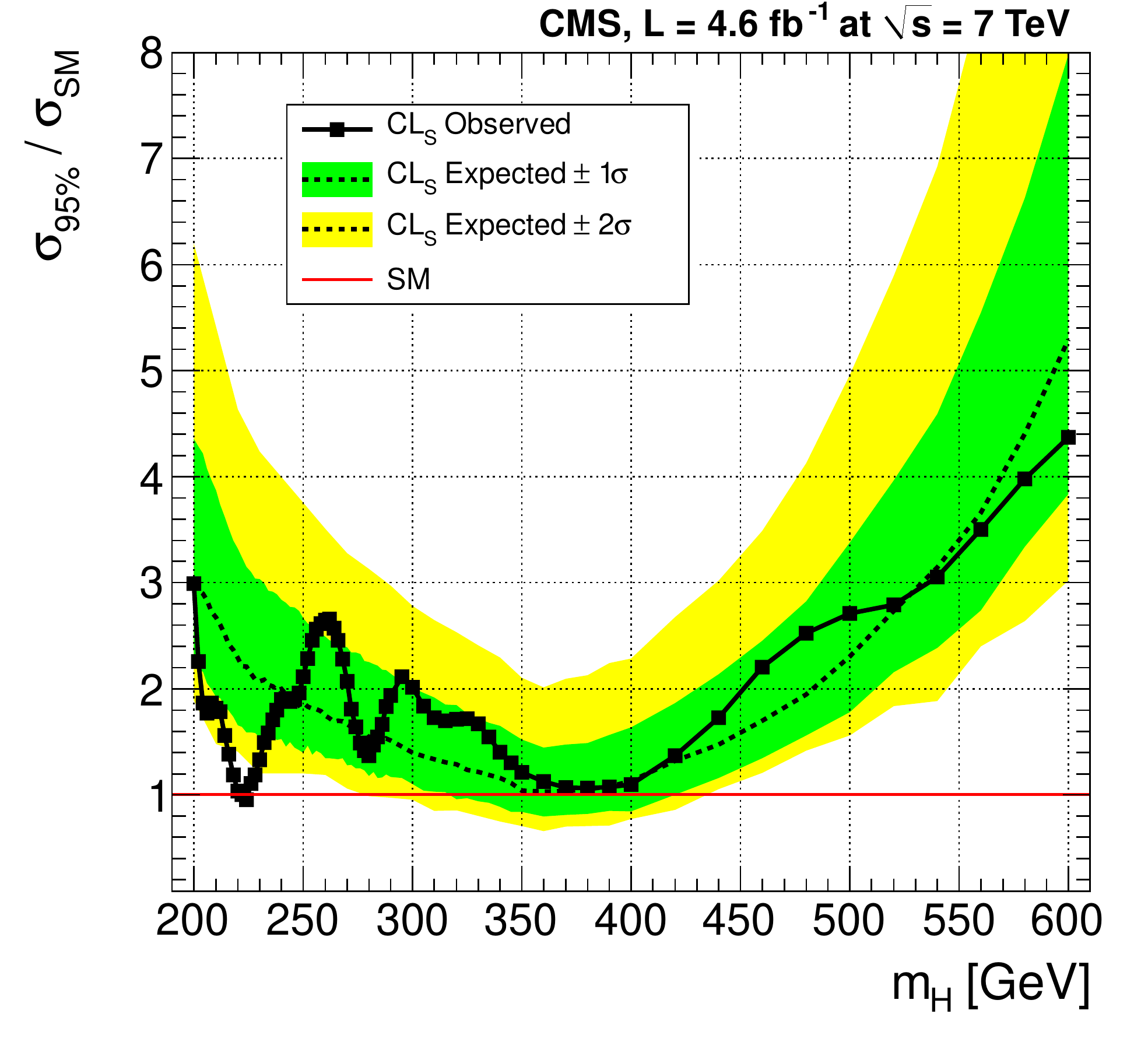}
}
\caption{
Observed (solid) and expected (dashed) 95\% CL upper limit
on the ratio of the production cross section to the SM expectation
for the Higgs boson obtained using the $\mathrm{CL_s}$ technique.
The 68\% (1$\sigma$) and 95\% (2$\sigma$) ranges of expectation
for the background-only model are also shown with green (darker) and
yellow (lighter) bands, respectively.  The solid line at 1 indicates the SM
expectation. Left: low-mass range, right: high-mass range.
}
\label{fig:plot_ul}
\end{center}
\begin{center}
\centerline{
\includegraphics[width=0.5\linewidth]{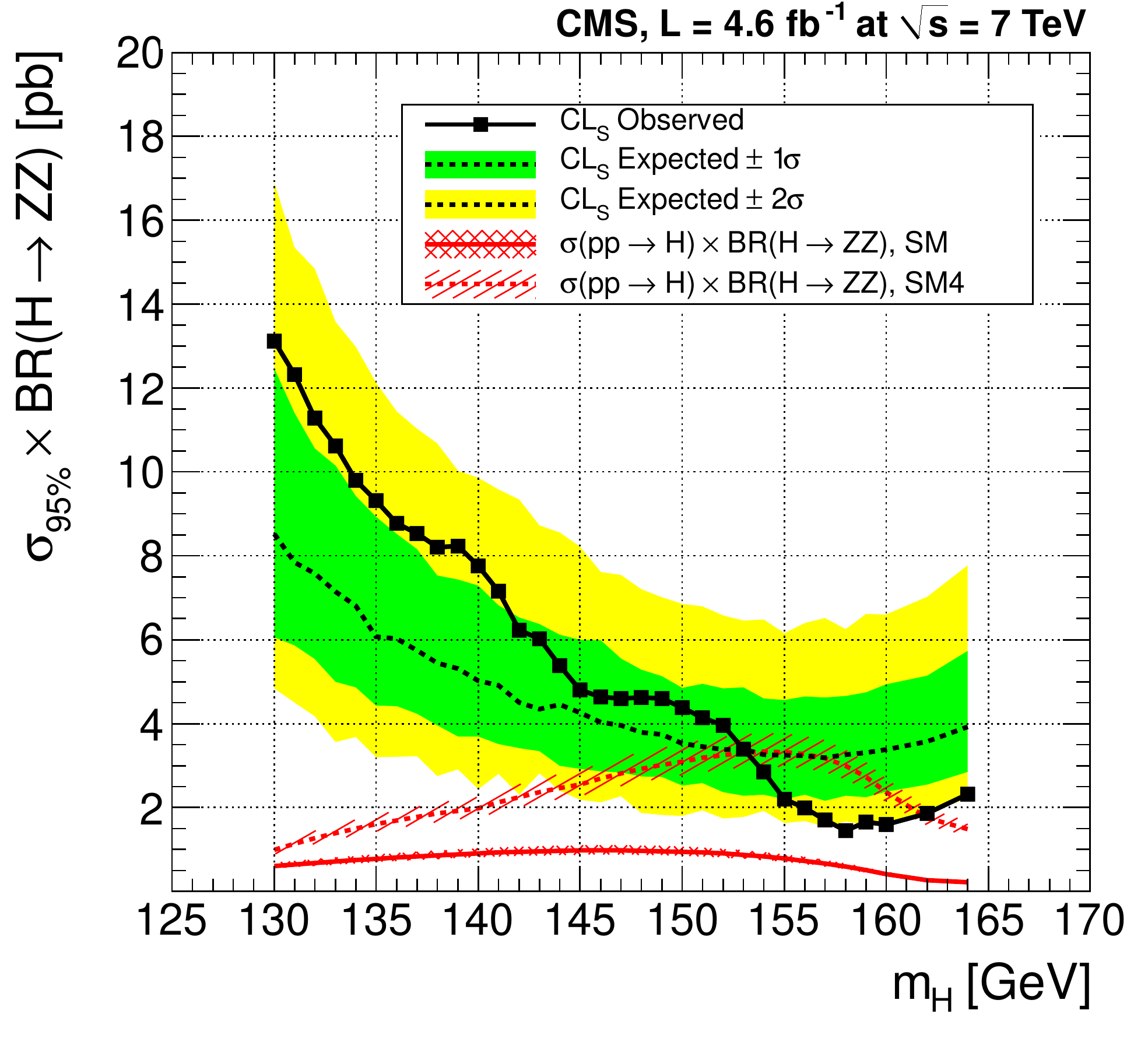}
\includegraphics[width=0.5\linewidth]{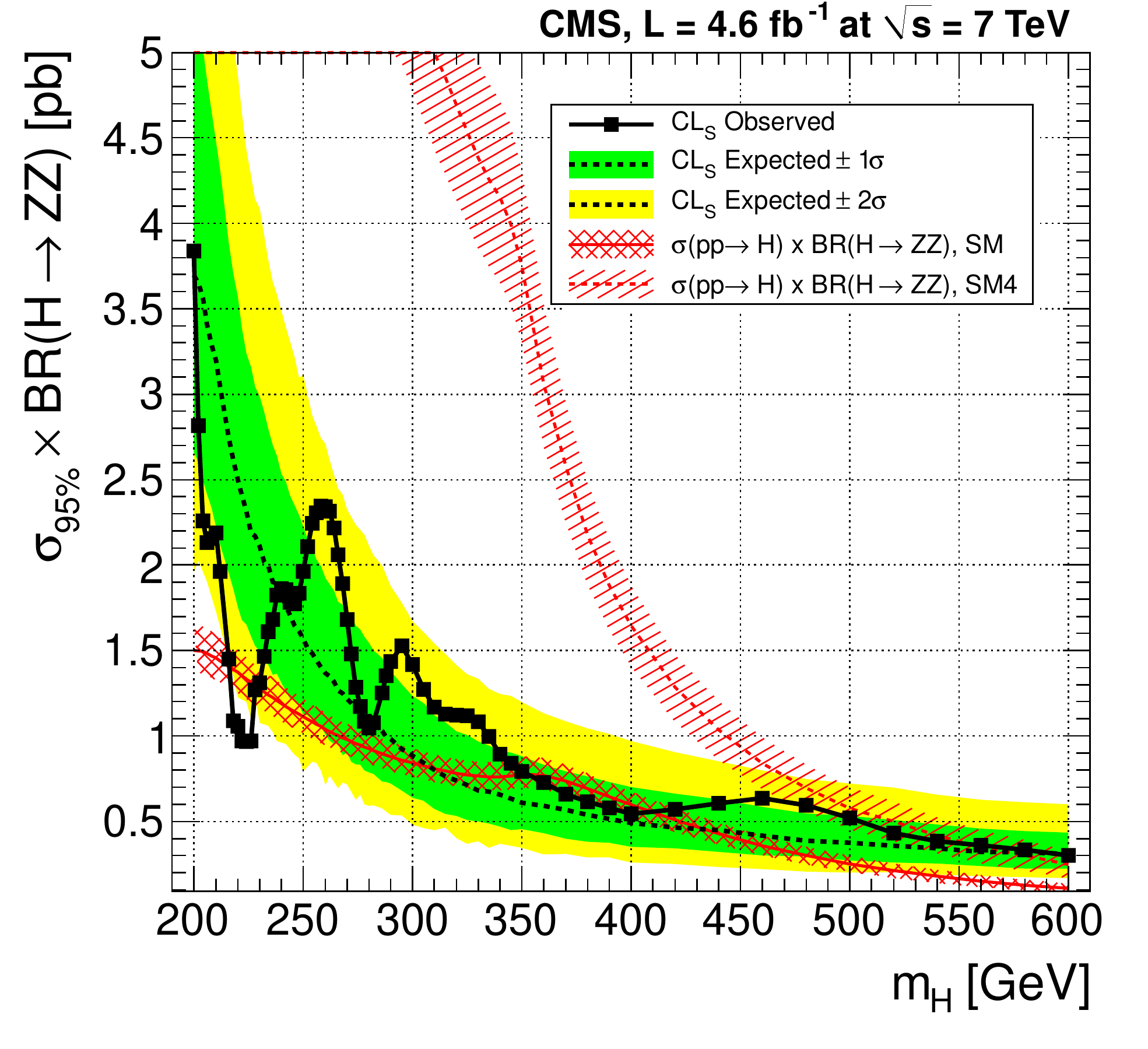}
}
\caption{
Observed (dashed) and expected (solid) 95\% CL upper
limit on the product of the  production cross section and
branching fraction for $\HZZ$ obtained with the $\mathrm{CL_s}$ technique.
The 68\% (1$\sigma$) and 95\% (2$\sigma$) ranges of expectation for the background-only model are
also shown with green (darker) and yellow (lighter) bands, respectively.
The expected product of the SM Higgs production cross section
and the branching fraction is shown as a red solid curve with
a band indicating theoretical uncertainties at 68\% CL.
The same expectation in the fourth-generation model is shown with a red dashed curve with
a band indicating theoretical uncertainties.
Left: low-mass range, right: high-mass range.
}
\label{fig:plot_sigma}
\end{center}
\end{figure}

No evidence for the Higgs boson is found and exclusion limits at 95\% CL on the ratio of the production
cross section for the Higgs boson to the SM expectation are presented in figure~\ref{fig:plot_ul}.
The observed limits are within expectation for the background-only model.
The significance of the only local deviation beyond the 95\% expectation range
around 225\GeV
is greatly reduced after taking into account the look-elsewhere effect~\cite{LEE},
for which the estimated trial factor is about 18 in the high-mass range.
Results obtained with the Bayesian approach yield very similar limits to those from $\mathrm{CL_s}$.

Limits on the SM production cross section times branching fraction for
$\HZZ$ are presented in figure~\ref{fig:plot_sigma}.
For comparison, expectations are shown for both the SM and for the SM4 model.
The ranges 154--161\GeV and 200--470\GeV
of SM4 Higgs mass hypotheses are excluded at 95\%~CL.
The exclusion limits in figure~\ref{fig:plot_ul} are also approaching the cross section
for the SM expectation for production of the Higgs boson.

\section{Summary}
\label{sec:summary}

A search for the SM Higgs boson decaying into two $\Zo$ bosons which subsequently
decay to two quark jets and two leptons, $\HZZllqq$, has been presented.
Data corresponding to an integrated luminosity of 4.6\fbinv
of proton-proton collisions at centre-of-mass energy of 7\TeV
have been collected and analyzed by the CMS Collaboration at the LHC.
No evidence for a SM-like Higgs boson has
been found and upper limits on the production cross section for the SM Higgs boson
have been set in the range of masses between
130 and 164\GeV, and between 200 and 600\GeV.
In this analysis we have also excluded at 95\%~CL a large range of Higgs boson mass hypotheses
in the model with a fourth generation of fermions having SM-like couplings.

\section*{Acknowledgements}

We wish to congratulate our colleagues in the CERN accelerator departments for the excellent performance of the LHC machine. We thank the technical and administrative staff at CERN and other CMS institutes, and acknowledge support from: FMSR (Austria); FNRS and FWO (Belgium); CNPq, CAPES, FAPERJ, and FAPESP (Brazil); MES (Bulgaria); CERN; CAS, MoST, and NSFC (China); COLCIENCIAS (Colombia); MSES (Croatia); RPF (Cyprus); MoER, SF0690030s09 and ERDF (Estonia); Academy of Finland, MEC, and HIP (Finland); CEA and CNRS/IN2P3 (France); BMBF, DFG, and HGF (Germany); GSRT (Greece); OTKA and NKTH (Hungary); DAE and DST (India); IPM (Iran); SFI (Ireland); INFN (Italy); NRF and WCU (Korea); LAS (Lithuania); CINVESTAV, CONACYT, SEP, and UASLP-FAI (Mexico); MSI (New Zealand); PAEC (Pakistan); MSHE and NSC (Poland); FCT (Portugal); JINR (Armenia, Belarus, Georgia, Ukraine, Uzbekistan); MON, RosAtom, RAS and RFBR (Russia); MSTD (Serbia); MICINN and CPAN (Spain); Swiss Funding Agencies (Switzerland); NSC (Taipei); TUBITAK and TAEK (Turkey); STFC (United Kingdom); DOE and NSF (USA).
Individuals have received support from the Marie-Curie programme and the European Research Council (European Union); the Leventis Foundation; the A. P. Sloan Foundation; the Alexander von Humboldt Foundation; the Belgian Federal Science Policy Office; the Fonds pour la Formation \`a la Recherche dans l'Industrie et dans l'Agriculture (FRIA-Belgium); the Agentschap voor Innovatie door Wetenschap en Technologie (IWT-Belgium); the Council of Science and Industrial Research, India; and the HOMING PLUS programme of Foundation for Polish Science, cofinanced from European Union, Regional Development Fund.

\bibliography{auto_generated}   

\cleardoublepage \appendix\section{The CMS Collaboration \label{app:collab}}\begin{sloppypar}\hyphenpenalty=5000\widowpenalty=500\clubpenalty=5000\textbf{Yerevan Physics Institute,  Yerevan,  Armenia}\\*[0pt]
S.~Chatrchyan, V.~Khachatryan, A.M.~Sirunyan, A.~Tumasyan
\vskip\cmsinstskip
\textbf{Institut f\"{u}r Hochenergiephysik der OeAW,  Wien,  Austria}\\*[0pt]
W.~Adam, T.~Bergauer, M.~Dragicevic, J.~Er\"{o}, C.~Fabjan, M.~Friedl, R.~Fr\"{u}hwirth, V.M.~Ghete, J.~Hammer\cmsAuthorMark{1}, M.~Hoch, N.~H\"{o}rmann, J.~Hrubec, M.~Jeitler, W.~Kiesenhofer, M.~Krammer, D.~Liko, I.~Mikulec, M.~Pernicka$^{\textrm{\dag}}$, B.~Rahbaran, C.~Rohringer, H.~Rohringer, R.~Sch\"{o}fbeck, J.~Strauss, A.~Taurok, F.~Teischinger, P.~Wagner, W.~Waltenberger, G.~Walzel, E.~Widl, C.-E.~Wulz
% ...
\vskip\cmsinstskip
\dag:~Deceased\\
1:~~Also at CERN, European Organization for Nuclear Research, Geneva, Switzerland\\
% ...

\end{sloppypar}
\end{document}